\documentclass[aps,prd,twocolumn,floatfix,superscriptaddress,showpacs,amsmath,notitlepage,nofootinbib]{revtex4-1}
\usepackage{graphicx}
\usepackage{bm}
\usepackage{slashed}
\usepackage{amsmath}
\usepackage{enumitem}
\usepackage{hhline}
\usepackage{multirow}
\usepackage[utf8]{inputenc} 
\usepackage[dvipsnames]{xcolor}
\usepackage[colorlinks,citecolor=blue]{hyperref}

\begin{document}
\title{Collective fast neutrino flavor conversions in a 1D box: Initial conditions and long-term evolution}

\author{Meng-Ru Wu}
\email{mwu@gate.sinica.edu.tw}
\affiliation{Institute of Physics, Academia Sinica, Taipei, 11529, Taiwan}
\affiliation{Institute of Astronomy and Astrophysics, Academia Sinica, Taipei, 10617, Taiwan}
\author{Manu George}
\email{manug@gate.sinica.edu.tw}
\affiliation{Institute of Physics, Academia Sinica, Taipei, 11529, Taiwan}
\author{Chun-Yu Lin}
\email{lincy@nchc.org.tw}
\affiliation{National Center for High-performance computing, National Applied Research Laboratories, Hsinchu Science Park, Hsinchu City
30076, Taiwan}
\author{Zewei Xiong}
\email{z.xiong@gsi.de}
\affiliation{GSI Helmholtzzentrum f\"ur Schwerionenforschung, Planckstra{\ss}e 1, 64291 Darmstadt, Germany}

\date{\today}

\begin{abstract}
We perform numerical simulations of fast collective neutrino flavor conversions in 
an one-dimensional box mimicking a system with the periodic boundary condition in one spatial direction and translation symmetry in the other two dimensions.
We evolve the system over several thousands of the characteristic timescale (inverse of the interaction strength) with different initial $\bar\nu_e$ to $\nu_e$ number density ratios and
different initial seed perturbations.
We find that small scale structures are formed 
due to the interaction of the flavor waves.
This results in nearly flavor depolarization in a certain neutrino phase space, when averaged over the entire box.
Specifically, systems with initially equal number of $\nu_e$ and $\bar\nu_e$ can reach full flavor depolarization for the entire neutrino electron lepton number ($\nu$ELN) angular spectra.
For systems with initially unequal $\nu_e$ 
and $\bar\nu_e$, flavor depolarization can only be reached in one side of the $\nu$ELN spectra,
dictated by the net neutrino $e-x$ lepton number
conservation.
Quantitatively small differences depending on the initial perturbations are also found when different perturbation seeds are applied.
Our numerical study here provides new insights for efforts aiming to include impact of fast flavor conversions in astrophysical simulations while calls for better analytical understanding accounting for the evolution of fast flavor conversions. 
\end{abstract}
\maketitle
\section{Introduction}\label{intro}

The discovery of the flavor oscillations of neutrinos by terrestrial experiments and astrophysical observations is one of the most exciting milestones in neutrino physics.
Ongoing and planned experimental projects are expected to further pin down the yet-unknown parameters in neutrino mixing -- the mass ordering and the \emph{CP}-violating phase -- while searching for potential signatures of physics beyond the Standard Model~\cite{Zyla:2020zbs}.

However, despite the success of the theory of neutrino mixing in explaining the majority of experimental data, one aspect that is poorly understood yet is how neutrinos oscillate in astrophysical environments dense in neutrinos.
Analytical and numerical works over the past decades have shown that in an environment where the self-interactions between neutrinos cannot be ignored, various collective phenomena can arise due to the nonlinear and strong coupling nature (in the flavor space) of the system; see e.g., \cite{Duan:2005cp,Duan:2006an,Hannestad:2006nj,Esteban-Pretel:2008ovd,Dasgupta:2009mg,Malkus:2012ts,Cherry:2012zw,Raffelt:2013rqa,Vlasenko:2013fja,Volpe:2013uxl,Wu:2015fga,Abbar:2015fwa,Sawyer:2015dsa,Izaguirre:2016gsx,Capozzi:2018clo,Abbar:2020ror,Xiong:2021dex,Johns:2021qby} and review articles~\cite{Duan:2010bg,Mirizzi:2015fva,Duan:2015cqa,Tamborra:2020cul}.
Exploratory works also demonstrated that such collective neutrino flavor oscillations may largely affect our understanding of important astrophysical events such as the core-collapse supernovae and the merger of two neutron stars~\cite{Duan:2010af,Wu:2014kaa,Sasaki:2017jry,Wu:2017drk,Wu:2017qpc,Stapleford:2019yqg,Xiong:2020ntn,George:2020veu,Li:2021vqj}.

Among these efforts, an important aspect that was recently pointed out is the potential occurrence of the ``fast'' neutrino flavor conversions~\cite{Sawyer:2015dsa}.
Fast flavor conversions happen when the angular distribution of the neutrino electron lepton numbers ($\nu$ELN) take both positive and negative values -- dubbed ``crossings''.
Under the two-flavor approximation in $\nu_e$--$\nu_x$ subspace where $\nu_x$ refers to a linear combination of $\nu_\mu$ and $\nu_\tau$, 
a $\nu$ELN crossing means that the effective differential neutrino $e-x$ number density
$d{\tilde n}_\nu/d\Omega$, where $\tilde n_{\nu}=n_{\nu_e}-n_{\bar\nu_e}-n_{\nu_x}+n_{\bar\nu_x}$ and $\Omega$ is the solid angle,
transits from positive to negative (or vice versa) in the angular phase space.
This has been confirmed by means of linear instability analysis, valid when the flavor conversion probabilities remain small, as well as numerical studies~\cite{Izaguirre:2016gsx,Dasgupta:2016dbv,Capozzi:2017gqd,Abbar:2018beu,Yi:2019hrp,Martin:2019gxb,Padilla-Gay:2020uxa,Johns:2020qsk,Bhattacharyya:2020dhu,Bhattacharyya:2020jpj,Morinaga:2021vmc,Richers:2021nbx,Zaizen:2021wwl,Kato:2021cjf}.
Meanwhile, searches for the conditions where fast conversions can develop using neutrino angular distributions provided by hydrodynamical simulations of supernovae and mergers were carried out extensively~\cite{Morinaga:2019wsv,Nagakura:2019sig,Johns:2019izj,DelfanAzari:2019tez,Abbar:2019zoq,Glas:2019ijo,Abbar:2020fcl,Johns:2021taz,Nagakura:2021hyb}.

In particular, Refs.~\cite{Martin:2019gxb} and \cite{Bhattacharyya:2020jpj} examined how fast flavor conversions develop in the nonlinear regime using numerical simulations in a one-dimensional (1D) box which possesses translation symmetry in the $x$ and $y$ directions while periodic in the $z$ direction.
Reference~\cite{Martin:2019gxb} showed that coherent and wavelike patterns in flavor space can develop with a point-source like perturbation.
On the other hand, Ref.~\cite{Bhattacharyya:2020jpj} found that the system can quickly settle into a state where flavor depolarization happens when averaged over the $z$ domain, for a major part of neutrino spectrum with either point-source like or random perturbation seeds.
These findings seem to contradict each other and require further examinations.

In this work, we systematically investigate the dependence of the outcome of fast neutrino flavor conversions in an 1D box on the initial condition of the perturbation seeds.
By adopting advanced numerical schemes, we evolve the systems to late times when quasi steady states are achieved. 
We also explore the dependence of the steady-state outcome on the initial $\bar\nu_e$ to $\nu_e$ number density ratio.

In Sec.~\ref{sec:model}, we describe our models, the initial and boundary conditions, and the adopted numerical schemes.
We also briefly review the notation of the neutrino polarization vectors.
In Sec.~\ref{sec:evo}, we discuss our numerical simulation results and the implications.
We summarize our main findings in Sec.~\ref{sec:summary}.
Throughout the paper, we adopt natural units and take $\hbar=c=1$.

\section{Models}\label{sec:model}

\subsection{Equation of motion}
\label{sec:eom}
We consider a simple neutrino-dense system which has translation symmetry in the $x$- and $y$- directions in space, the same as in Refs.~\cite{Martin:2019gxb,Bhattacharyya:2020dhu,Bhattacharyya:2020jpj}.
For simplicity, initially we assume a pure system consisting of only $\nu_e$ and $\bar\nu_e$
(before setting small flavor perturbations; see below).
Focusing on fast flavor conversions, we neglect the momentum changing collisions and only keep the neutrino--neutrino forward scattering contributions~\cite{Fuller:1987,Pantaleone:1992eq,Sigl:1992fn} in the effective Hamiltonian, i.e., we drop the terms originating from vacuum neutrino mixing and neutrino--matter forward scattering contributions.

Assuming the differential neutrino angular distribution $dn_{\nu_e (\bar\nu_e)}/d\Omega$ is uniform in $z$ and taking the two-flavor approximation, 
the equation of motion, which governs the evolution of the normalized neutrino density matrices (see below)
$\varrho$ (for neutrinos) and $\bar\varrho$ (for antineutrinos) is given as follows\footnote{When only considering the neutrino--neutrino self-interaction term, one can redefine the antineutrino density matrix by  $\bar\rho\rightarrow -\sigma_y^\dagger \bar\rho \sigma_y$ such that neutrinos and antineutrinos are treated in equal footing~\cite{Duan:2005cp}.
However, here we choose to treat them separately for the purpose of consistency with follow-up works that may include other terms in the Hamiltonian and the collisions.}:
\begin{subequations}\label{eq:eom}
\begin{align}
& \frac{\partial}{\partial t}\varrho(t,z,v_z)+
v_z\frac{\partial}{\partial z}\varrho(t,z,v_z)
=-i[H(t,z,v_z),\varrho(t,z,v_z)],\\
& \frac{\partial}{\partial t}\bar\varrho(t,z,v_z)+
v_z\frac{\partial}{\partial z}\bar\varrho(t,z,v_z)
=-i[\bar H(t,z,v_z),\bar\varrho(t,z,v_z)].
\end{align}
\end{subequations}
with
\begin{equation}\label{eq:rho}
\varrho(t,z,v_z)=
\left [
\begin{array}{cc}
\varrho_{ee} & \varrho_{ex}\\
\varrho_{ex}^* & \varrho_{xx} \\
\end{array}
\right ],~
\bar\varrho(t,z,v_z)=
\left [
\begin{array}{cc}
\bar\varrho_{ee} & \bar\varrho_{ex}\\
\bar\varrho_{ex}^* & \bar\varrho_{xx} \\
\end{array}
\right ],
\end{equation}
in the flavor basis. 
Since we omit the vacuum mixing contribution, the flavor evolution of $\rho$ and $\bar\rho$ do not depend on the neutrino energy.

In Eq.~\eqref{eq:eom}, the Hamiltonian $H$ and $\bar H$ can be explicitly written down as:
\begin{subequations}\label{eq-hnu}
\begin{align}
 H(t,z,v_z)= &
~ \mu \int_{-1}^1 dv_z' (1-v_z v_z')\times\\
& [g_\nu(v_z')\varrho(t,z,v_z')-
\alpha g_{\bar\nu}(v_z')\bar\varrho^*(t,z,v_z')],\nonumber\\
 \bar H(t,z,v_z)=&
-
\mu \int_{-1}^1 dv_z' (1-v_z v_z')\times \\
& [g_\nu(v_z')\varrho^*(t,z,v_z')-\alpha g_{\bar\nu}(v_z')\bar\varrho(t,z,v_z')],\nonumber
\end{align}
\end{subequations}
where $\mu=\sqrt{2}G_F n_{\nu_e}$ with $G_F$ being the Fermi coupling constant. 
The asymmetry parameter $\alpha=n_{\bar\nu_e}/n_{\nu_e}$ quantifies the ratio between the number density of $\bar\nu_e$ and $\nu_e$.
The angular distribution function $g_{\nu(\bar\nu)}(v_z)$ relates to the physical neutrino phase-space distribution function $f_{\nu_e(\bar\nu_e)}$ without any flavor perturbations as
\begin{equation}
g_{\nu(\bar\nu)}(v_z)=\frac{1}{4\pi^2 n_{\nu_e (\bar\nu_e)}}\int dE_\nu E_\nu^2 
f_{\nu_e (\bar\nu_e)},
\end{equation}
where $E_\nu$ is the energy of a neutrino. Note that the azimuthal symmetry in phase space of $f_{\nu_e(\bar\nu_e)}$ with respect to the $z$ direction is implicitly taken.
In this notation, $\rho_{ee}=\bar\rho_{ee}=1$
for our system without any initial flavor perturbations while all other matrix elements are zero.

Since $\mu$ is the only dimensional quantity in Eq.~\eqref{eq:eom}, we define $\mu\equiv 1$ and express $t$ and $z$ in dimensionless form (in the units of $\mu^{-1}$) hereafter.

\subsection{Initial and boundary conditions}
\label{sec:ibc}

\begin{figure}[t]
    \centering
    \includegraphics[width=1.0\linewidth]{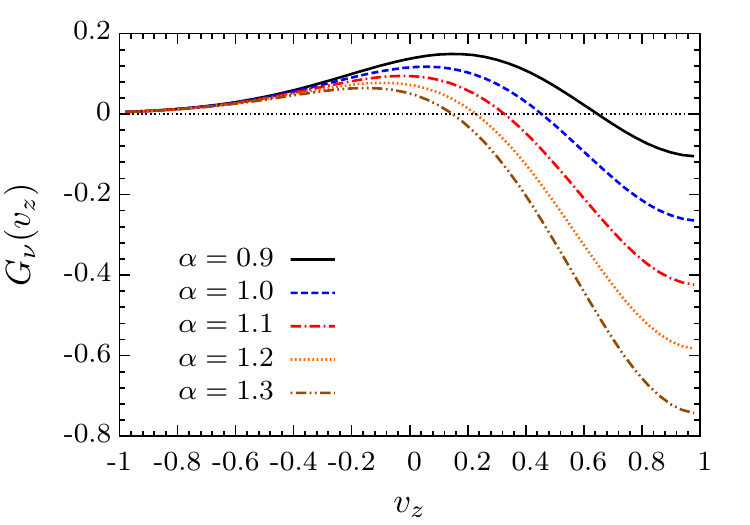}
    \caption{The initial $\nu$ELN angular distribution $G(v_z)$ for systems with different $\bar\nu_e$ to $\nu_e$ number density ratios $\alpha$.
    The $\nu$ELN crossing where $G(v_z)=0$ shifts to smaller $v_z$ with larger $\alpha$. 
    }
    \label{fig:Gv}
\end{figure}

We simulate the flavor evolution of neutrinos 
inside a box of size $L_z=1200$ in $z\in [-600,600]$ with the periodic boundary condition.
Similar to Ref.~\cite{Martin:2019gxb}, we parametrize $g_{\nu(\bar\nu)}$ as
\begin{equation}\label{eq:gv}
    g_{\nu (\bar\nu)}(v_z)\propto\exp[-(v_z-1)^2/(2\sigma_{\nu(\bar\nu)}^2)],
\end{equation}
with normalization condition $\int_{-1}^1 dv_z g_{\nu(\bar\nu)}(v_z)=1$.
Throughout the paper, we fix $\sigma_\nu=0.6$ and $\sigma_{\bar\nu}=0.5$, such that $\bar\nu_e$ are more forward-peaked than $\nu_e$.
For the asymmetry parameter $\alpha$, 
we take $\alpha=0.9, 1.0$, $1.1$, $1.2$, and $1.3$.
Figure~\ref{fig:Gv} shows the corresponding $\nu$ELN distributions $G_\nu(v_z)\equiv g_\nu(v_z)-\alpha g_{\bar\nu}(v_z)$.
With increasing value of $\alpha$, the $\nu$ELN crossing where $G_\nu(v_{z,c})=0$ occurs at smaller $v_{z,c}$ with $|G_\nu(v_z)|$ being larger (smaller) for $v_{z}>v_{z,c}$ ($v_{z}<v_{z,c}$).
The values of $v_{z,c}$ for $\alpha=0.9, 1.0, 1.1, 1.2$, and $1.3$ are 0.65, 0.45, 0.33, 0.23, and 0.15, correspondingly.

For a system without the vacuum Hamiltonian, it requires a small perturbation in $\rho_{ex}$ ($\bar\rho_{ex}$) to trigger the flavor instabilities.
For simplicity, we assume that the perturbations are independent of $v_z$.
Specifically, we use the following description for our initial condition at $t=0$ as:
\begin{subequations}
\begin{align}
 & \varrho_{ee}(z,v_z) = \bar\varrho_{ee}(z,v_z) = (1+\sqrt{1-\epsilon^2 (z)})/2, \\
 & \varrho_{xx}(z,v_z) = \bar\varrho_{xx}(z,v_z) =(1-\sqrt{1-\epsilon^2 (z)})/2, \\
 & \varrho_{ex}(z,v_z)  =\bar\varrho_{ex}(z,v_z) = \epsilon (z)/2.
\end{align}
\end{subequations}
For $\epsilon(z)$, we explore the following two different choices:
\begin{enumerate}
\item Point-source like perturbations centered at $z=0$: $\epsilon(z)=\epsilon_0\exp[-z^2/50]$.
\item Random perturbations: $\epsilon(z)$ is randomly assigned by a real value between $0$ and $\epsilon_0$.
\end{enumerate}
We take $\epsilon_0=10^{-6}$ throughout the paper.

\subsection{Numerical methods}
\label{sec:num}
We discretize $z$ and $v_z$ with $N_z$ and $N_{v_z}$ grids, and evolve $\varrho$ and $\bar\varrho$ defined at the grid centers.
In evaluating the advection terms $\partial \varrho/{\partial  z}$ and $\partial \bar\varrho/{\partial  z}$, we use two different methods to check for consistency.
The first method uses the forth-order (central) finite difference method with artificial dissipation using the third-order Kreiss-Oliger formulation~\cite{KO3}.
For the second method, we use finite volume method plus the seventh order weighted essentially nonoscillatory (WENO) scheme~\cite{Shu1998,Shu2003}.
For time evolution, we use fourth-order Runge-Kutta method.
The numerical implementation details will be reported in a separate publication together with the public release of our simulation code \texttt{COSE$\nu$}~\cite{COSEnu}.

Using the above initial and boundary conditions, we perform simulations with fiducial numerical parameters of $N_Z=12000$ and $N_{v_z}=200$, with a fixed time step size $\Delta t=C_{\rm CFL}\Delta z$, where $\Delta z = L_z/N_Z$ and the Courant–Friedrichs–Lewy number $C_{\rm CFL}=0.4$.
In Appendixes~\ref{app:res} and \ref{app:method}, we show the comparison of results obtained with different resolutions and with two different numerical methods for advection.
For the rest of the paper, all results are obtained using the finite volume method with seventh order WENO scheme, which provides better numerical accuracy given the same resolution.

\subsection{Polarization vectors}
Before we discuss our results, let us define
the neutrino and antineutrino polarization vectors $\mathbf{P}$ and $\mathbf{\bar P}$ that are often used in literature.
The three components of the polarization vector $\mathbf{P}$ are defined as 
\begin{subequations}\label{eq:Pnu}
\begin{align}
P_1 &=  2{\rm Re}(\varrho_{ex}),\\
P_2 &=  -2{\rm Im}(\varrho_{ex}),\\
P_3 &=  (\varrho_{ee}-\varrho_{xx}),
\end{align}
\end{subequations}
which satisfy $\rho\equiv (\mathbf{P}\cdot \bm{\sigma}+P_0\mathcal{I})/2$, where $\sigma_i$ are the Pauli matrices and $\mathcal{I}$ is the identity matrix.
For antineutrinos, we have 
\begin{subequations}\label{eq:Pnub}
\begin{align}
{\bar P}_1 &=  2{\rm Re}(\bar\varrho_{ex}),\\
{\bar P}_2 &=  2{\rm Im}(\bar\varrho_{ex}),\\
{\bar P}_3 &=  (\bar\varrho_{ee}-\bar\varrho_{xx}).
\end{align}
\end{subequations}

With this definition, the equation of motion for $\mathbf{P}$ has the same form as for $\mathbf{\bar P}$, 
and we can treat neutrinos and antineutrinos in equal footing using the $\nu$ELN spectrum $G_\nu$ defined in Sec.~\ref{sec:ibc} to account for the contributions from both neutrinos and antineutrinos in the Hamiltonian.
Dropping the bar for antineutrinos, 
we have:
\begin{equation}\label{eq:eompvec}
\left(\frac{\partial}{\partial t}+v_z\frac{\partial }{\partial z}\right)
\mathbf{P}(t,z,v_z) = \mathbf{H}(t,z,v_z)\times \mathbf{P}(t,z,v_z),
\end{equation}
where $\mathbf{H}(t,z,v_z)=\int_{-1}^1 dv'_z G_\nu(v'_z) \mathbf{P}(t,z,v'_z) (1-v_z v'_z)$.
Clearly, in the absence of collisions, the length of the polarization vectors are unity for all $z$ and $v_z$, given that our initial condition has $|\mathbf{P}|=1$ uniformly in $z$.
Another physical quantity that is conserved globally is the net neutrino $e-x$ lepton number in the entire simulation domain~\cite{Bhattacharyya:2020dhu,Bhattacharyya:2020jpj}:
\begin{equation}\label{eq:netlep}
L_{e-x} \propto \int_{-L_z/2}^{L_z/2} dz  \int_{-1}^1 dv_z G_\nu(v_z) P_3(t,z,v_z).
\end{equation}
Note that locally this net lepton number can be transferred from one location to another due to the advection.
However, globally $L_{e-x}$ must be a conserved quantity in time for the entire system.

\begin{figure*}[t]
    \centering
    \includegraphics[width=1.0\linewidth]{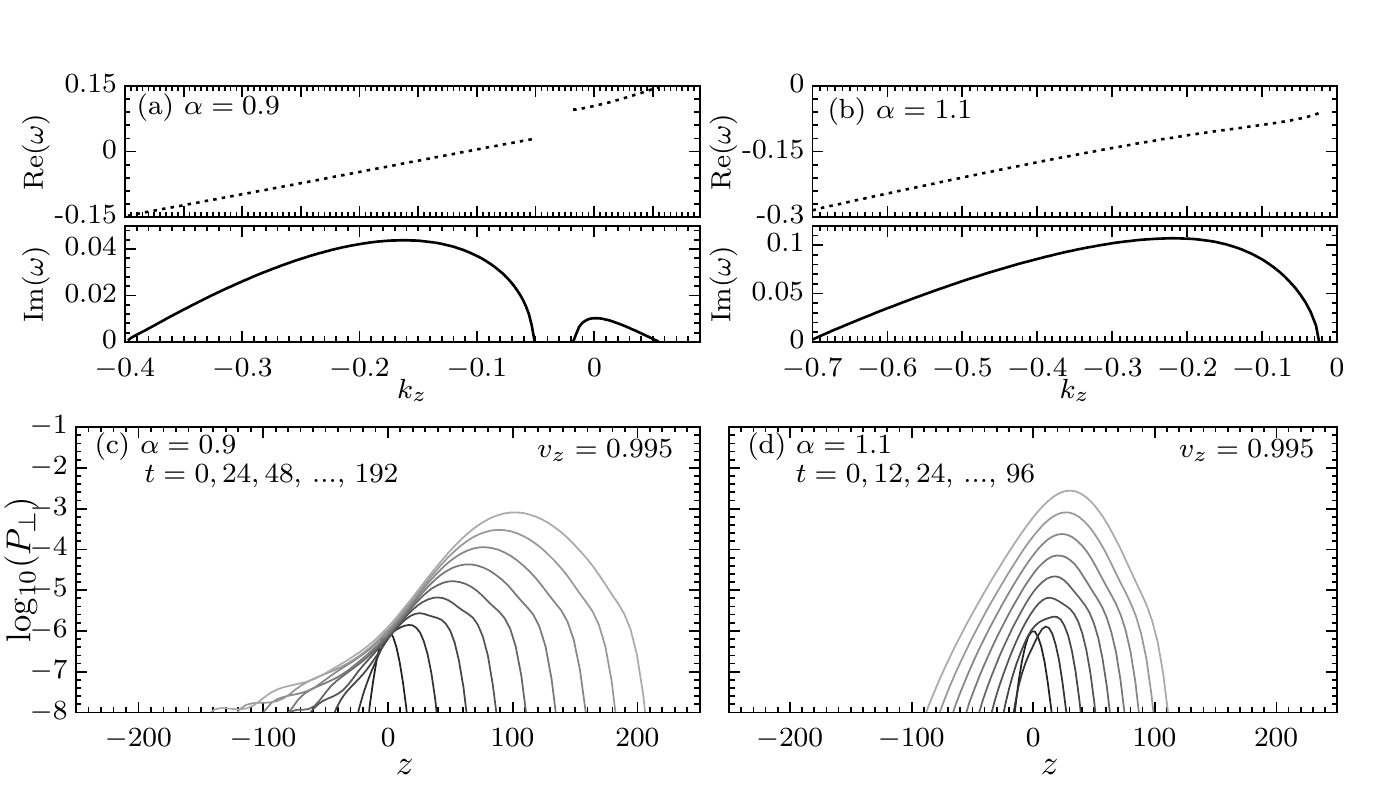}
    \caption{Panels (a) and (b): The dispersion relation of $\omega(k_z)$ for unstable solutions containing nonzero $\rm{Im}(\omega)$ for $\alpha=0.9$ and $1.1$. 
    Panels (c) and (d): Time evolution of the perpendicular components of the neutrino polarization vectors $P_\perp(z)$ in the linear regime ($|P_\perp|\ll 1$) for $v_z=0.995$. 
    Darker curves correspond to earlier times.
    The same point-source like flavor perturbations are applied for both cases with $\alpha=0.9$ [panel (c)] and $\alpha=1.1$ [panel (d)].
    }
    \label{fig:pperp_linear}
\end{figure*}

\section{Results: flavor evolution of the system}
\label{sec:res}

In this section, we focus on results obtained with $\alpha=0.9$ and $\alpha=1.1$.
In Sec.~\ref{sec:linear}, we examine how the point-source like perturbations in the off diagonal elements of the density matrices grow in the linear regime and compare their evolution with detailed analysis using the dispersion relation.
In Secs.~\ref{sec:pointpert} and \ref{sec:randpert}, we further discuss how flavor evolution proceeds in the nonlinear regime for cases with point-source like perturbations and random perturbations, respectively.

\subsection{Linear regime}

\label{sec:linear}

In the regime where $|\rho_{ex}|\ll 1$ and $|\bar\rho_{ex}|\ll 1$, 
the evolution of the system can be qualitatively understood by performing the standard technique of linearized instability analysis~\cite{Banerjee:2011fj,Izaguirre:2016gsx}.  
By inserting $\rho_{ex} \propto \exp[-i(\omega t - k_z z)]$
(same for $\bar\rho_{ex}$) and keeping terms in the lowest order of perturbations,
Eq.~\eqref{eq:eom} leads to a dispersion relation of the collective mode of the system $\omega(k_z)$~\cite{Izaguirre:2016gsx,Capozzi:2017gqd,Yi:2019hrp}.
The complex branch of $\omega$ for real $k_z$ signifies the existence of flavor instabilities such that small off diagonal perturbations can grow exponentially in time as the system evolves.
Note that here we only include the solution that preserves the the axial symmetry.
For the symmetry-breaking solutions, 
we omit them here because our simulation setup does not allow symmetry-breaking solutions.

In Fig.~\ref{fig:pperp_linear}, we show in panels (a) and (b) the dispersion relation $\omega(k_z)$
for systems with $\nu$ELN asymmetry parameter $\alpha=0.9$ and $1.1$.
For $\alpha=0.9$, there are two regions in $k_z$ where complex $\omega$ exist. 
The maximal value of Im$(\omega_{\rm max})\simeq 0.044$ locates at $k_{z,{\rm max}}\simeq -0.165$. 
At the same $k_{z,{\rm max}}$, $(d\omega/d k_z)_{\rm max}\simeq 0.515$.
For $\alpha=1.1$ where the $\nu$ELN crossing is deeper in larger $v_z$ (see Fig.~\ref{fig:Gv}), Im$(\omega_{\rm max})\simeq 0.107$ at $k_{z,{\rm max}}\simeq -0.225$. The corresponding $(d\omega/d k_z)_{\rm max}\simeq 0.278$.

In the bottom panels [(c) and (d)] of Fig.~\ref{fig:pperp_linear}, 
we show the evolution of $|P_\perp(z)|=\sqrt{P_1^2+P_2^2}=2|\rho_{ex}|$ 
for $v_z=0.995$ in the liner regime for the same $\nu$ELNs with point-source like perturbations.
Clearly, an initial Gaussian packet of $P_\perp$ grows exponentially with time and the peak position of $P_\perp$ moves toward positive $z$
direction for both the cases.
The growth rate of the maximal value of $P_{\perp,{\rm max}}$ and the velocity  $dz(P_{\perp,{\rm max}})/dt$ both agree well with
the values of Im$(\omega_{\rm max})$ and $(d\omega/d k_z)_{\rm max}$ obtained in the dispersion relation above.
The growth of perturbations using $\alpha=1.1$ is indeed much faster than that with $\alpha=0.9$.
Moreover, the $P_\perp(z)$ with $\alpha=1.1$ spreads over both positive and negative $z$ directions. 
For the case with $\alpha=0.9$, although the width of $P_\perp(z)$ also increases with time, 
it mainly drifts to the positive $z$ direction without spreading over the negative $z$ direction.

\begin{figure*}[t]
    \centering
    \includegraphics[width=1.05\linewidth]{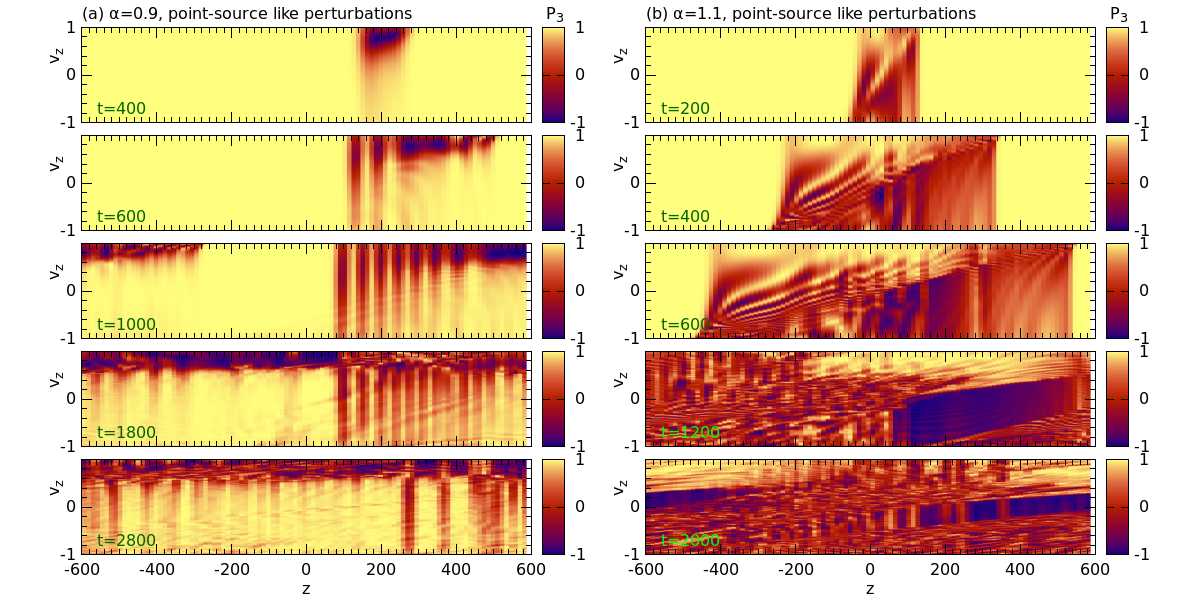}
    \caption{Snapshots of $P_3(z,v_z)$ at different simulation times for the same systems shown in Fig.~\ref{fig:pperp_linear} when the initial perturbations have grown to the nonlinear regime.
    Flavor waves with coherent structures develop and propagate.
    Small scale structures form when flavor waves interact and breaks the coherent structure.
    }
    \label{fig:p3evol_ps}
\end{figure*}

\subsection{Nonlinear regime: Point-source like perturbations}
\label{sec:pointpert}

When the perturbations grow to the nonlinear regime, wavelike oscillatory features develop~\cite{Martin:2019gxb}.
In Fig.~\ref{fig:p3evol_ps}, we show the snapshots of $P_3(z,v_z)$ at different times
for $\alpha=0.9$ and $1.1$ using the same point-source like perturbations as in Sec.~\ref{sec:linear}.
For the case of $\alpha=0.9$, the flavor evolution behavior of the system is in general similar to those reported in Ref.~\cite{Martin:2019gxb}.
Flavor waves develop and mainly propagate toward the positive $z$ direction.
This is consistent with the growth of perturbations in the linear regime discussed earlier.
An interesting feature shown here is that although the flavor oscillations initially can affect all $v_z$ modes, illustrated by the vertical stripes in the upper two subpanels in Fig.~\ref{fig:p3evol_ps}(a), 
this effect diminishes when time proceeds and flavor conversions are roughly confined in $v_z\gtrsim v_{z,c}\simeq 0.65$.

Another interesting feature is when the forward-traveling wave front interacts with the slowly backward propagating part after $t\gtrsim 1200$, it pushes the whole pattern to drift toward positive $z$ direction.
Meanwhile, this interaction breaks the coherent wavelike pattern. 
Substructures in smaller scale develop such that the orientation of the neutrino polarization vectors varies rapidly in $z$ for $v_z\gtrsim v_{z,c}$.
Consequently, although at each location $z$, neutrinos with different $v_z\gtrsim v_{z,c}$ still have $|\mathbf{P}|=1$, the ``average polarization vector'' over the $z$ domain can shrink due to their misalignment. 
Such a flavor state was referred as ``flavor depolarization'' in Refs.~\cite{Bhattacharyya:2020dhu,Bhattacharyya:2020jpj} and we will discuss its behavior in more detail in the next section.
For $v_z\lesssim v_{z,c}$, most neutrinos remain unaffected.

For $\alpha=1.1$ shown in panel (b), flavor conversions quickly develop toward both the positive and negative $z$ directions and produces coherent and wavelike structure, once again consistent with that indicated by the growth of perturbations in the linear regime.
Similarly, when the forward and backward propagating modes interact after $t\gtrsim 665$, smaller structures develop and cause a major part of $v_z$ reaching closer to flavor depolarization.
One important difference from the previous $\alpha=0.9$ case is now flavor depolarization happens mostly in $v_z\lesssim v_{z,c}\simeq 0.45$. 
We will also discuss this feature and its consequences in Sec.~\ref{sec:evo_vmode}.

For the other $\nu$ELN spectra with $\alpha=1.0$, $1.2$, and $1.3$, the behaviors are qualitatively similar to that with $\alpha=1.1$.
Full simulation animations are available at \url{https://mengruwuu.github.io/COSEv1dbox/}.

\subsection{Nonlinear regime: Random perturbations}
\label{sec:randpert}

\begin{figure*}[t]
    \centering
    \includegraphics[width=1.05\linewidth]{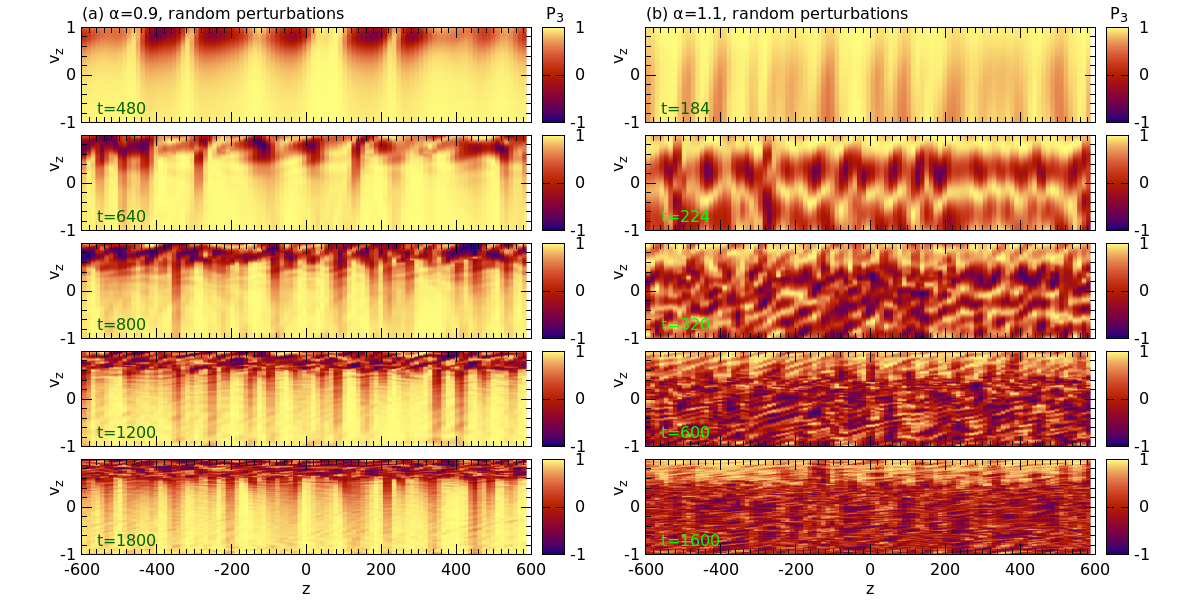}
    \caption{Snapshots of $P_3(z,v_z)$ for systems with the same $\nu$ELN shown in Fig.~\ref{fig:p3evol_ps} but with different seed perturbations.
    With random perturbations, interaction of flavor waves happen much earlier. 
    No large-scale coherent structures can be formed when compared to Fig.~\ref{fig:p3evol_ps}, which adopts point-source like perturbations.
    }
    \label{fig:p3evol_rd}
\end{figure*}

Now, let us look at how the flavor systems evolve when we adopt different seed of random perturbations.
We show again $P_3(z,v_z)$ at different time snapshots in Fig.~\ref{fig:p3evol_rd} for $\alpha=0.9$ and $1.1$.
With random perturbations, some locations have larger $P_\perp$ initially such that flavor conversions develop faster around those locations (see the top subpanels).
Similar to what discussed in 
Sec.~\ref{sec:pointpert} for single point-source like perturbations, flavor waves can initially develop independently now at different locations.
The flavor waves transported in space 
interact and break the coherent pattern to form small-scale structures. 
Thus, with perturbations everywhere in space, the interaction of flavor waves happen at much earlier times and no large-scale coherent structure can be formed, differently from cases with point-source like perturbations.
The systems can then reach closer to flavor depolarization in $v_z\gtrsim v_{z,c}$ ($v_z\lesssim v_{z,c}$) for $\alpha=0.9$ ($1.1$) within a much shorter amount of time.

Once again, for the other $\nu$ELN spectra with $\alpha=1.0$, $1.2$, and $1.3$, the behaviors are qualitatively similar to that with $\alpha=1.1$.
Full simulation animations are also available at \url{https://mengruwuu.github.io/COSEv1dbox/}.

\section{Results: evolution of averaged quantities}
\label{sec:evo}
In the previous Sec.~\ref{sec:res}, we have discussed how fast flavor conversions of neutrinos can develop in space and time with different seed perturbations.
In this section, we further examine the 
time evolution of relevant quantities averaged over $z$ and/or $v_z$ in  Sec.~\ref{sec:evo_overall} and Sec.~\ref{sec:evo_vmode}.

\subsection{Overall flavor conversion probability in the box}
\label{sec:evo_overall}

\begin{figure*}[t]
    \centering
    \includegraphics[width=1.0\linewidth]{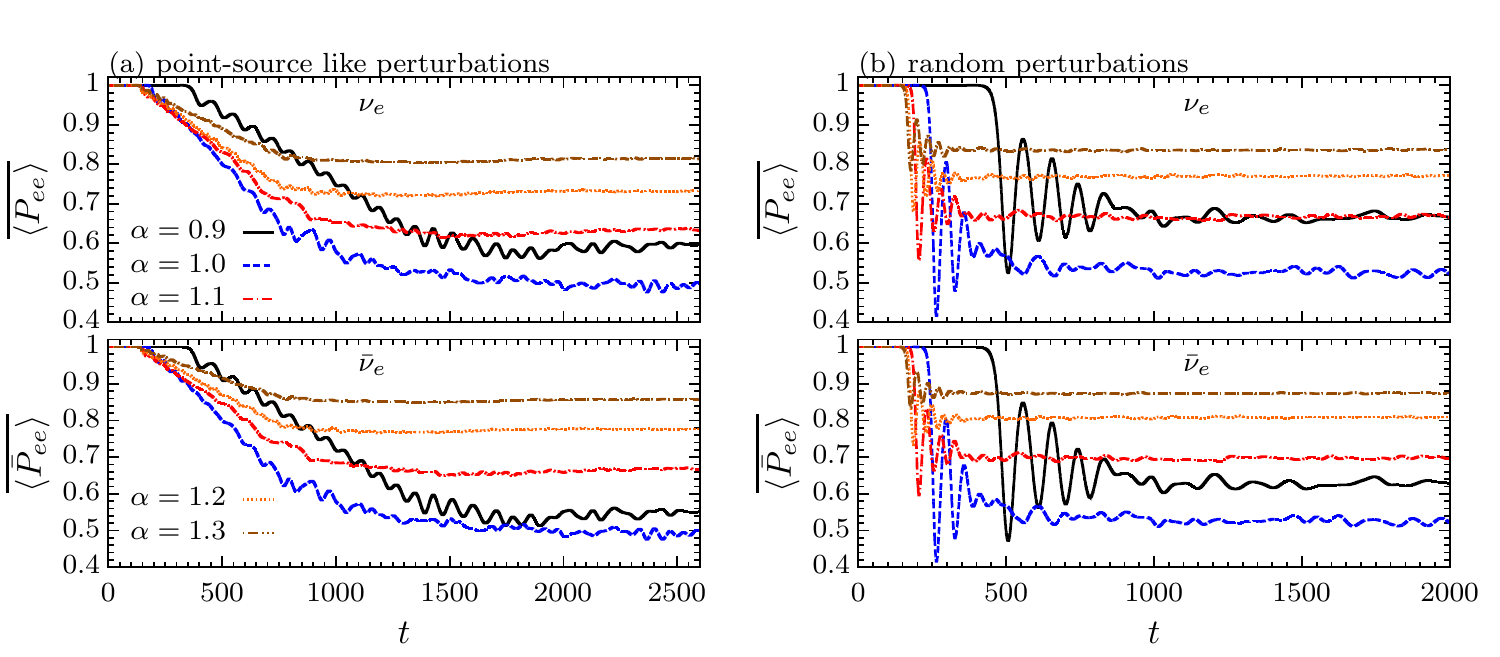}
    \caption{Time evolution of the flavor survival probabilities [see Eq.~\eqref{eq:avgPall}] for $\nu_e$ (upper subpanels) and $\bar\nu_e$ (lower subpanels) averaged over $z$ and $v_z$ 
    for systems with point-source like perturbations [panel (a)] and random perturbations [panel (b)] with different $\bar\nu_e$ to $\nu_e$ number density ratios $\alpha$. See text in Sec.~\ref{sec:evo_overall} for details.}
    \label{fig:avgprob}
\end{figure*}

We define the overall flavor survival probabilities of neutrinos and antineutrinos, $\overline{\langle P_{ee}\rangle}$ and $\overline{\langle {\bar P}_{ee}\rangle}$ in the simulation domain by
\begin{subequations}\label{eq:avgPall}
\begin{align}
\overline{\langle P_{ee}\rangle} 
= \int dz dv_z \varrho_{ee}(z,v_z) g_\nu(v_z) \Bigg/ \int dz dv_z g_\nu(v_z),\\
\overline{\langle {\bar P}_{ee}\rangle} 
= \int dz dv_z \bar\varrho_{ee}(z,v_z) g_{\bar\nu}(v_z) \Bigg/ \int dz dv_z g_{\bar\nu}(v_z),
\end{align}
\end{subequations}
where the brackets and overline denote averaging over $z$ and $v_z$, respectively.
The integration limits over $z$ and $v_z$ are from $-L_z/2$ to $L_z/2$ and from $-1$ to $1$, respectively.

Fig.~\ref{fig:avgprob} shows the time evolution of $\overline{\langle P_{ee}\rangle}$ and $\overline{\langle {\bar P}_{ee}\rangle}$ for all five different $\nu$ELN spectra that we considered (see Fig.~\ref{fig:Gv}).
The cases with point-source like perturbations and random perturbations are shown in panels (a) and (b), respectively.
First, we see that cases with larger $\alpha$ reach the final asymptotic states earlier in time for both point-source like and random perturbations.
Meanwhile, all cases with random perturbations reach the asymptotic states much earlier than those with point-source like perturbations as discussed in Sec.~\ref{sec:randpert}.

Second, systems with $\alpha=1$ (equal number of neutrinos and antineutrinos) 
have final asymptotic values of $\overline{\langle P_{ee}\rangle}_f\simeq \overline{\langle {\bar P}_{ee}\rangle}_f \simeq 0.5$, i.e., full flavor depolarization is almost reached for the entire system.
For systems that are more asymmetric, 
the asymptotic $\overline{\langle P_{ee}\rangle}_f$ and $ \overline{\langle {\bar P}_{ee}\rangle}_f$ 
are larger, i.e., on average less $\nu_e$
and $\bar\nu_e$ are converted.
Comparing $\overline{\langle P_{ee}\rangle}_f$ to $ \overline{\langle {\bar P}_{ee}\rangle}_f$ for a given $\alpha$, one finds that cases with $\alpha > 1$ have larger values of asymptotic $\overline{\langle {\bar P}_{ee}\rangle}_f$ than those of $\overline{\langle P_{ee}\rangle}_f$ (vice versa for $\alpha<1$).
This is related to the fact that 
for $\alpha>1$ ($\alpha<1$), neutrinos with $v_z\lesssim v_{z,c}$ ($v_z \gtrsim v_{z,c}$) experience more flavor conversions. 
Since for our adopted neutrino angular spectra, $g_{\bar\nu}(v_z)$ is more forward peaked in positive $v_z$ than $g_{\nu}(v_z)$, more flavor conversions in larger $v_z$ naturally lead to smaller 
$\overline{\langle {\bar P}_{ee}\rangle}_f$ than $\overline{\langle P_{ee}\rangle}_f$.

\begin{figure}[t]
    \centering
    \includegraphics[width=1.0\linewidth]{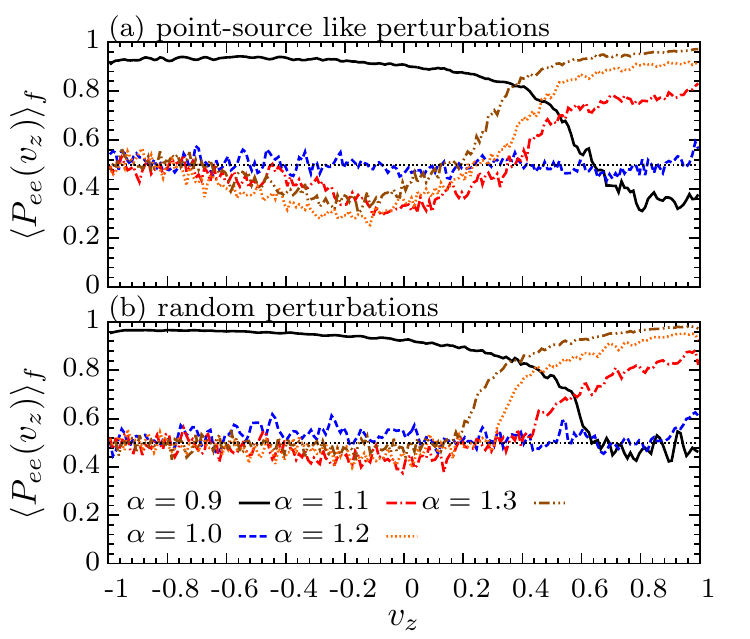}
    \caption{The final flavor survival probabilities $\langle P_{ee}(v_z)\rangle_f$ averaged over $z$ [Eq.~\eqref{eq:Peevz}]
    for the same systems shown in Fig.~\ref{fig:avgprob}.
    Nearly flavor depolarization can be reached in one side of the $\nu$ELN spectra for cases with $\alpha\neq 1$. 
    For $\alpha=1$, the entire $\nu$ELN spectra can achieve nearly flavor depolarization.
    }
    \label{fig:fv}
\end{figure}

Comparing the final values of
$\overline{\langle P_{ee}\rangle}_f$
and $\overline{\langle {\bar P}_{ee}\rangle}_f$
for cases with point-source like and random perturbations, the point-source like ones generally lead to a slightly smaller 
$\overline{\langle P_{ee}\rangle}_f$
and $\overline{\langle {\bar P}_{ee}\rangle}_f$ than the random ones.
This is due to the reason discussed in Sec.~\ref{sec:pointpert}: although the interaction of the flavor waves lead to states close to flavor depolarization, 
for cases with point-source like perturbations, some large-scale coherent structures can remain until the end of the simulations.
This can also be seen in Fig.~\ref{fig:fv}, which shows the spatially averaged flavor survival probabilities 
$\langle P_{ee}(v_z)\rangle_f$
as functions of $v_z$ at the end of our simulations, where $\langle P_{ee}(v_z)\rangle$ is defined by
\begin{subequations}\label{eq:Peevz}
\begin{align}
\langle P_{ee}(v_z)\rangle
= \int dz \varrho_{ee}(z,v_z) \Bigg/ \int dz.
\end{align}
\end{subequations}
Note that $\langle {\bar P}_{ee}(v_z)\rangle$ that can be similarly defined are equal to $\langle P_{ee}(v_z)\rangle$, when only neutrino-neutrino self-interaction terms are included here.

Figure~\ref{fig:fv} clearly shows several interesting features.
First, for cases with either point-source like or random perturbations, those with $\alpha>1$ ($\alpha<1$) lead to 
nearly full flavor depolarization ($\langle P_{ee}(v_z)\rangle_f\simeq 0.5$) for velocity modes $v_z<v_{z,c}$ ($v_z>v_{z,c}$).
For velocity modes in the other side of the $\nu$ELN, nearly flavor equipartition cannot be achieved and $\langle P_{ee}(v_z)\rangle_f$ gradually increase from $\simeq 0.5$ to larger values for larger (smaller) $v_z$ for
$\alpha>1$ ($\alpha<1$).
For the case with $\alpha=1$, i.e., symmetric in neutrinos and antineutrinos, full flavor depolarization for the entire $\nu$ELN can be achieved.
Comparing results with different perturbations, one sees that due to the remaining large-scale coherent structures (see e.g., Fig.~\ref{fig:p3evol_ps}), systems with point-source like perturbations give rise to smaller 
$\langle P_{ee}(v_z)\rangle_f\simeq 0.4$ in some regions in $v_z<v_{z,c}$ ($v_z>v_{z,c}$) for $\alpha>1$ ($\alpha<1$) than those with random perturbations. 

Based on our simulation results, we make a conjecture as follows.
For a system with the periodic boundary condition in one spatial direction (while having perfect translation symmetry in the other two directions) where interaction of flavor waves can happen, 
the system can evolve to an asymptotic state where nearly flavor equipartition can be reached for velocity modes in one side of the $\nu$ELN, when averaged over space.
In other words, when averaged over space, the $\nu$ELN crossing nearly vanishes.
For such a scenario to happen, the net neutrino $e-x$ lepton number conservation $L_{e-x}$ [see Eq.~\eqref{eq:netlep}] would
enforce that nearly flavor depolarization can only happen for velocity modes in the shallower side of the $\nu$ELN.

An additional remark is: comparing Fig.~\ref{fig:avgprob} and Fig.~\ref{fig:fv}, we also point out that although Fig.~\ref{fig:fv} seems to suggest that a narrower range of velocity modes in $\nu$ELN distribution reaches nearly flavor depolarization for $\alpha=0.9$ than those with $\alpha>1$,
on average, more neutrinos and antineutrinos are being converted.
Once again, this is because the spectra $g_\nu(v_z)$ and $g_{\bar\nu}(v_z)$ 
are more forward peaked.
This means that although a system with $\alpha<1$ may appears to be affected in less phase space volume in $v_z$, 
flavor conversions there can actually have a larger impact on physical processes that are flavor dependent in realistic astrophysical environments.

\subsection{Evolution of different velocity modes}
\label{sec:evo_vmode}

\begin{figure}[t]
    \centering
    \includegraphics[width=1.0\linewidth]{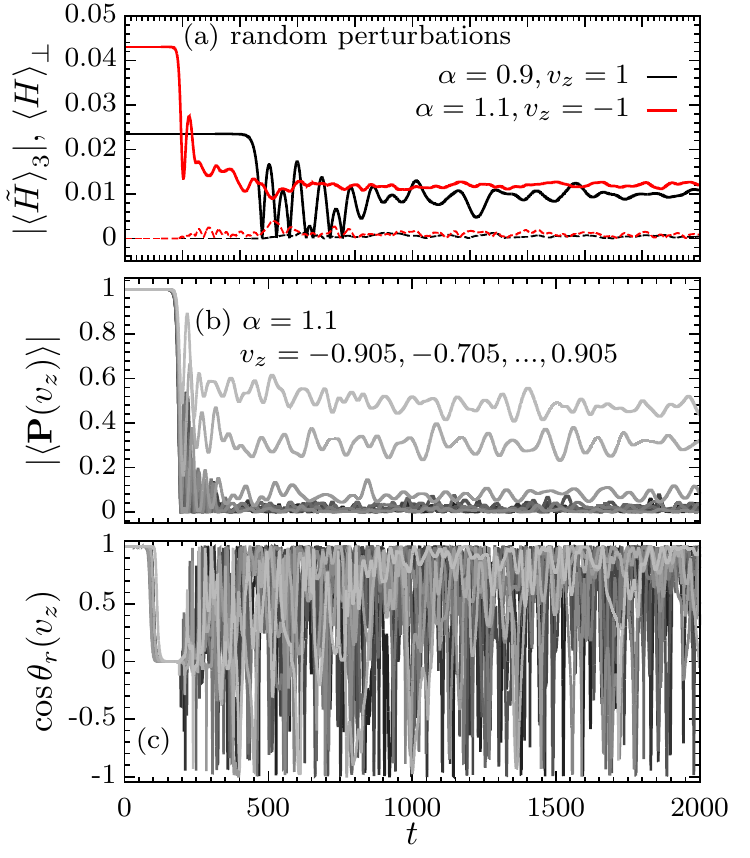}
    \caption{Time evolution of different spatially averaged quantities.
    See text in Sec.~\ref{sec:evo_vmode} for details.
    }
    \label{fig:vmode}
\end{figure}

Refs.~\cite{Bhattacharyya:2020dhu,Bhattacharyya:2020jpj} proposed that the behavior of the system can be understood by examining the time evolution of  
$\langle \mathbf{P}(v_z)\rangle$.
From Eq.~\eqref{eq:eompvec}, 
one obtains
\begin{equation}\label{eq:Pv}
\frac{d}{dt}\langle \mathbf{P}(v_z,t) \rangle
=\langle \mathbf{M}_0 \times \mathbf{P}(v_z,t) \rangle -
v_z \langle \mathbf{M}_1 (t) \times \mathbf{P}(v_z,t) \rangle,
\end{equation}
where $\mathbf{M}_i(z,t)\equiv \int dv_z L_i(v_z) G_\nu(v_z) \mathbf{P}(z,v_z,t)$
with $L_i(v_z)$ the Legendre Polynomials.
The authors of Refs.~\cite{Bhattacharyya:2020dhu,Bhattacharyya:2020jpj}
argued that the spatial averaging of the cross products in
Eq.~\eqref{eq:Pv} can be approximated as the cross products of two vectors that are spatially averaged separately:
\begin{equation}\label{eq:Pv2}
\frac{d}{dt}\langle \mathbf{P}(v_z,t) \rangle
\simeq [\langle \mathbf{M}_0 \rangle
- v_z \langle \mathbf{M}_1 (t)\rangle]
\times \langle \mathbf{P}(v_z,t) \rangle. 
\end{equation}
By doing so, the time evolution of 
$\langle \mathbf{P}(v_z,t) \rangle$
may be understood as a vector precessing around an effective Hamiltonian vector
$\langle\mathbf{H}(v_z)\rangle=\langle \mathbf{M}_0 \rangle
- v_z \langle \mathbf{M}_1 (t)\rangle$.
Moreover, the depolarization of 
$\langle \mathbf{P}(v_z,t) \rangle$
happens when the $\langle H\rangle_\perp=\sqrt{\langle H\rangle_1^2+\langle H\rangle_2^2}$ roughly exceeds $|\langle {\tilde H}\rangle_3|$~\cite{Bhattacharyya:2020jpj}, where $\langle {\tilde H}\rangle_3$
differs from $\langle H\rangle_3$ by a factor related to $\langle \mathbf{M}_2 \rangle$ when going to a co-rotating frame (see Ref.~\cite{Bhattacharyya:2020dhu} for details).

We carefully examine the validity of these claims.
In Fig.~\ref{fig:vmode}, we show in panel (a) the time evolution of 
$|\langle \tilde{H} \rangle_3|$ and $\langle H_\perp \rangle$ for $\alpha=0.9$ of $v_z=1$ and $\alpha=1.1$ of $v_z=-1$ using random
perturbations as examples. 
Despite these modes undergo nearly flavor depolarization, Fig.~\ref{fig:vmode}(a) shows that
the values of $\langle H_\perp \rangle$ remain much smaller than $|\langle \tilde{H} \rangle_3|$ nearly during the whole time, in contrast to what discussed in Ref.~\cite{Bhattacharyya:2020jpj}.

In panels (b) and (c) of Fig.~\ref{fig:vmode}, we show the evolution of $|\langle \mathbf{P}(v_z) \rangle|$ and the relative angles $\theta_r$ between the two vectors on the right-hand sides of Eqs.~\eqref{eq:Pv} and \eqref{eq:Pv2} for ten $v_z$ modes for the case with $\alpha=1.1$ with random perturbations.
Clearly, for modes with $v_z\lesssim v_{z,c}$ that reach flavor depolarization (darker colors), i.e., 
$|\langle \mathbf{P}(v_z) \rangle|\rightarrow 0$, 
their $\cos\theta_r$ oscillate rapidly between -1 and 1.
This indicates that Eq.~\eqref{eq:Pv2} is in fact not a good approximation of Eq.~\eqref{eq:Pv}.

In sum, our findings here suggest that the picture which explains the late-time behavior and the depolarization of the system proposed in Refs.~\cite{Bhattacharyya:2020dhu,Bhattacharyya:2020jpj} using the spatially averaged polarization vectors needs to be revisited.

\section{Summary and discussions}
\label{sec:summary}

We have performed long-term numerical simulations of fast collective neutrino flavor conversions
for systems with translation symmetry in two spatial ($x$ and $y$) directions using a newly developed code \texttt{COSE$\nu$}.
Our numerical calculations were conducted based on the finite volume method with seventh order WENO scheme and the finite difference method with Kreiss-Oliger dissipation. 
Both methods showed good capability of numerical error suppression which allows simulations being carried to longer than a few thousand characteristic timescale of the system.
Assuming uniform $\nu$ELN distributions in $z$ and taking
periodic boundary conditions,
we have investigated how flavor conversions happen with different 
$\nu$ELN spectra, controlled by the initial $\bar\nu_e$ to $\nu_e$ asymmetric ratio parameter $\alpha$,
as well as how the results depend on the
chosen flavor perturbations.
Our main findings can be summarized as follows.

First, we found that for systems with point-source like initial perturbations, flavor waves with coherent structures can develop and propagate.
With periodic boundary conditions,
these flavor waves can then interact and break
into smaller scale structures.
When adopting perturbation seed randomly placed in $z$, 
the flavor waves originated from different locations can interact much faster 
such that no coherent structure can be formed.

The interactions of the flavor waves lead the system to a final state where 
part of the velocity space ($v_z$) is close to 
flavor depolarization when averaging over the space.
For any asymmetric system with $\alpha\neq 1$, 
only one side of the $\nu$ELN spectrum relative to the $\nu$ELN crossing point
can reach close to an averaged flavor depolarization while neutrinos in the other side of $\nu$ELN experience less flavor conversions, as constrained by the
conservation of the net neutrino $e-x$ number.
Specifically, for $\alpha>1$ ($\alpha<1$), nearly flavor depolarization can be reached for $v_z\lesssim v_{z,c}$ ($v_z\gtrsim v_{z,c}$) when the antineutrinos angular distribution are more forward peaked than that of neutrinos.
On the other hand, for systems with $\alpha=1$, the entire neutrino spectra can reach close to averaged flavor depolarization.
This phenomenon is qualitatively similar for systems with either point-source like or random perturbations.
Quantitatively, the developed large-scale coherent pattern in cases with point-source like perturbations 
allow part of the velocity space to have
on average more flavor conversions than
perfect flavor depolarization.

Comparing our results with those reported in Refs.~\cite{Martin:2019gxb,Bhattacharyya:2020dhu,Bhattacharyya:2020jpj}, 
our results with point-source like perturbations agree with \cite{Martin:2019gxb}, which, however, only evolves the system for a shorter period of time without allowing flavor waves to interact. 
On the other hand, Refs.~\cite{Bhattacharyya:2020dhu,Bhattacharyya:2020jpj} obtained results nearly independent of whether the initial perturbations are being 
point-source like or random. 
In fact, behavior of random perturbations emerged in their
simulations using point-source like perturbations, different from what we 
obtained here.
One potential reason is that Refs.~\cite{Bhattacharyya:2020dhu,Bhattacharyya:2020jpj}
used fast Fourier transform to evaluate the derivative terms, 
which might artificially generate errors of random nature in the spatial domain.
Moreover, we have tried to verify the mechanisms proposed in Refs.~\cite{Bhattacharyya:2020dhu,Bhattacharyya:2020jpj} in explaining the occurrence of the flavor depolarization. 
However, our results do not support the proposed mechanisms and suggest that better understanding is needed.

Practically, our numerical findings may provide insights for efforts which attempt to include the impact of fast neutrino flavor conversions in hydrodynamical simulations; e.g.,~\cite{Li:2021vqj}.
For instance, if the adopted neutrino
transport scheme can provide the antineutrino-to-neutrino asymmetry ratio and the crossing points in the $\nu$ELN spectrum, partial flavor depolarization to one end of the $\nu$ELN together with the net $e-x$ neutrino lepton number 
may be applied to the neutrino distribution functions at where $\nu$ELN crossings are identified.

Needless to say, there are still several improvements to be made in future.
For example, our simulations were performed in reduced dimensions.
How the symmetry-breaking solutions may develop and affect our conclusions need to be further examined.
In our simulations, we have completely omitted the vacuum and the matter terms in the Hamiltonian, as well as the collision terms.
Adding these terms and incorporating full treatment with three neutrino flavors may introduce new effects recently investigated in works without advection~\cite{Capozzi:2020kge,Shalgar:2020xns,Shalgar:2020wcx,Martin:2021xyl}.
Full inclusion of them are to be implemented in future.
Last but not least, the potential impact of the many-body nature of the problem remain to be further elucidated~\cite{Rrapaj:2019pxz,Cervia:2019res,Roggero:2021asb,Roggero:2021fyo}.

\begin{acknowledgments}
We thank Geng-Yu Liu and Herlik Wibowo for useful discussions as well as Basudeb Dasgupta, Soumya Bhattacharyya, and Irene Tamborra for valuable feedback.
M.-R.~W. and M.~G. acknowledge supports from the Ministry of Science and Technology, Taiwan under Grants No.~109-2112-M-001-004, No.~110-2112-M-001 -050, and
the Academia Sinica under Project No.~AS-CDA-109-M11.
M.-R.~W. also acknowledge supports from the Physics Division, National Center for Theoretical Sciences, Taiwan.
M.-R.~W. and M.~G. appreciate the computing resources provided by the Academia Sinica Grid-computing Center.
C.-Y.~L. thank the National Center for High-performance Computing (NCHC) for providing computational and storage resources.
Z.~X. acknowledge supports from the European Research Council (ERC) under the European Union's Horizon 2020 research and innovation programme (ERC Advanced Grant KILONOVA No.~885281).
\end{acknowledgments}

\appendix
\section{Convergence tests}\label{app:res}
\begin{figure}[t]
    \centering
    \includegraphics[width=1.0\linewidth]{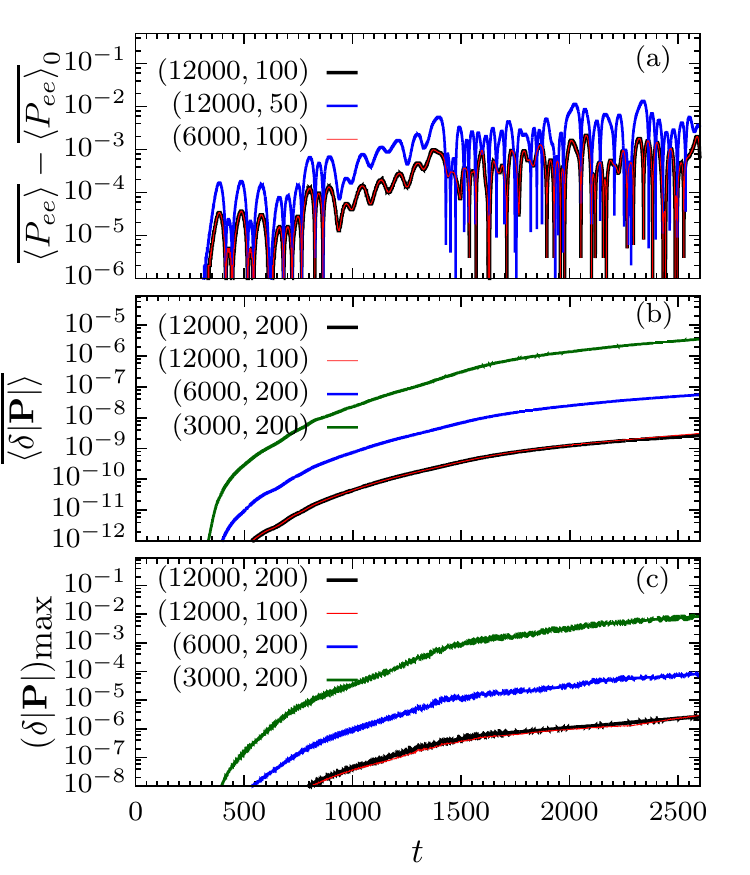}
    \caption{Panel (a): Difference in $\overline{\langle P_{ee} \rangle}$ between runs using different $(N_z,N_{v_z})$ and $\overline{\langle P_{ee} \rangle}_0$ obtained with $(N_z,N_{v_z})=(12000,200)$ with
    $\alpha=0.9$ and point-source like perturbations.
    Panels (b) and (c): Evolution of $\overline{\langle \delta |\mathbf{P}|\rangle }$ and $(\delta |\mathbf{P}|)_{\rm max}$ with different simulation resolutions.
    }
    \label{fig:res}
\end{figure}

In this appendix, we quantify errors associated with the
resolution of our numerical simulations. 
As described in the main text, our fiducial resolution 
is with $N_z=12000$ and $N_{v_z}=200$.
In panel (a) of Fig.~\ref{fig:res}, we
compare the absolute difference of 
$\overline{\langle P_{ee} \rangle}$ between runs with different resolutions and our fiducial case, 
taking $\alpha=0.9$ with point-source like perturbations.
It shows that the difference of $\overline{\langle P_{ee} \rangle}$ amounts to $\sim 10^{-3}$ and $\sim 10^{-2}$,
when we reduce $N_{v_z}$ by a factor of $2$ and $4$, respectively while keeping $N_z$ being the same.
For runs with fixed $N_{v_z}=200$ but with reduced $N_z=6000$ and $3000$, the relative differences are smaller than $10^{-6}$
and are thus not shown in the figure.
This can also be inferred by looking at the nearly identical 
black and red curves, which represent the differences
with $(N_z,N_{v_z})=(12000,100)$ and $(6000,100)$, respectively.

However, this does not mean that we can adopt a much reduced resolution in $N_z$.
In panels (b) and (c) of Fig.~\ref{fig:res},
we show the deviation of the length of the polarization vector from unity, averaged spatially and spectrally over $g_\nu(v_z)$:
\begin{equation}
\overline{\langle \delta |\mathbf{P}|}\rangle=\int dz dv_z g_\nu(v_z) (|\mathbf{P}(z,v_z)|-1) \Bigg/ \int dz dv_z g_\nu(v_z),
\end{equation}
and the maximal deviation of the length of the polarization 
vectors in the entire simulation domain $(\delta |\mathbf{P}|)_{\rm max}$.
Both panels clearly show that the errors are mostly depending on the resolution in $N_z$.
For our fiducial case with $(N_z,N_{v_z})=(12000,200)$,
the associated errors in $\overline{\langle \delta |\mathbf{P}|\rangle }$
and $(\delta |\mathbf{P}|)_{\rm max}$ are smaller than 
$10^{-8}$ and $10^{-5}$, respectively, for $\alpha=0.9$ with 
point-source like perturbations.
For all of our simulations with different $\alpha$ and different perturbation seeds, 
we obtain $\overline{\langle \delta |\mathbf{P}|}\rangle < 10^{-4}$.
For $(\delta |\mathbf{P}|)_{\rm max}$, all cases have $(\delta |\mathbf{P}|)_{\rm max}<10^{-2}$ except for $\alpha=1.2$ and $1.3$ with point-source like perturbations, for which their $(\delta |\mathbf{P}|)_{\rm max}$ reach $\mathcal{O}(1)$ by the end of the simulation.
We note, however, that demanding $(\delta |\mathbf{P}|)_{\rm max}\ll 1$ may not be a practical convergence criterion because $(\delta |\mathbf{P}|)_{\rm max}$ is usually associated with the grids with largest $|v_z|$, which has minor contribution to the Hamiltonian and the averaged quantities.
In fact, we have compared the errors in the averaged quantities for all cases with lowered resolutions ($N_z=6000$) and confirmed that all the averaged quantities agree within $5\times 10^{-3}$.

\section{Results using different numerical schemes}\label{app:method}

\begin{figure}[t]
    \centering
    \includegraphics[width=1.0\linewidth]{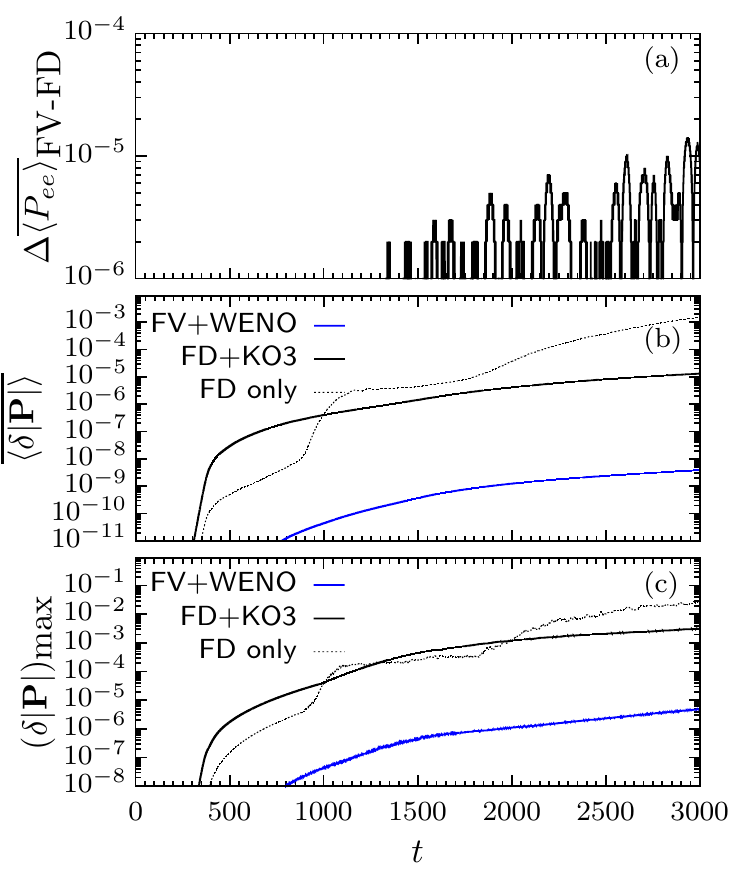}
    \caption{Comparison of the same quantities as those in Fig.~\ref{fig:res} for simulations with $\alpha=0.9$ and point-source like perturbations, using the finite volume method with seventh order WENO scheme (FV+WENO) and the finite difference method supplied by the third-order Kreiss-Oliger dissipation term (FD+KO3).
    Note that in panels (b) and (c), we show additionally the quantities derived using the finite difference method only without dissipation (FD only).
    }
    \label{fig:method}
\end{figure}

In this appendix, we compare our results obtained with
two different numerical methods described in Sec.~\ref{sec:num}.
Panel (a) of Fig.~\ref{fig:method} shows the difference of
$\overline{\langle P_{ee} \rangle}$ 
between the run using the finite volume method with seventh order 
WENO scheme (FV+WENO) and that with the finite difference method supplied by the third-order Kreiss-Oliger dissipation term (FD+KO3),
for $\alpha=0.9$ with point-source like perturbations and our fiducial resolution of $(N_z,N_{v_z})=(12000,200)$.
One sees that the differences are smaller than $10^{-5}$ throughout the whole time.
Panels (b) and (c) of the same figure show once again $\overline{\langle \delta |\mathbf{P}|\rangle }$
and $(\delta |\mathbf{P}|)_{\rm max}$.
Here we see that although both quantities remain much smaller than one with either the FV+WENO or the FD+KO3 scheme, the FV+WENO scheme gives rise to errors much smaller than those using FD+KO3 scheme.
In addition, we show values of $\overline{\langle \delta |\mathbf{P}|\rangle }$
and $(\delta |\mathbf{P}|)_{\rm max}$ using simply the finite difference method without applying any error suppression mechanisms (FD only) in panels (b) and (c) of Fig.~\ref{fig:method}. 
It shows clearly that both our FV+WENO and FD+KO3 schemes help suppress numerical errors in $\overline{\langle \delta |\mathbf{P}|\rangle }$ by more than two orders of magnitudes for this particular parameter set.

%


\begin{thebibliography}{70}%
\makeatletter
\providecommand \@ifxundefined [1]{%
 \@ifx{#1\undefined}
}%
\providecommand \@ifnum [1]{%
 \ifnum #1\expandafter \@firstoftwo
 \else \expandafter \@secondoftwo
 \fi
}%
\providecommand \@ifx [1]{%
 \ifx #1\expandafter \@firstoftwo
 \else \expandafter \@secondoftwo
 \fi
}%
\providecommand \natexlab [1]{#1}%
\providecommand \enquote  [1]{``#1''}%
\providecommand \bibnamefont  [1]{#1}%
\providecommand \bibfnamefont [1]{#1}%
\providecommand \citenamefont [1]{#1}%
\providecommand \href@noop [0]{\@secondoftwo}%
\providecommand \href [0]{\begingroup \@sanitize@url \@href}%
\providecommand \@href[1]{\@@startlink{#1}\@@href}%
\providecommand \@@href[1]{\endgroup#1\@@endlink}%
\providecommand \@sanitize@url [0]{\catcode `\\12\catcode `\$12\catcode
  `\&12\catcode `\#12\catcode `\^12\catcode `\_12\catcode `\%12\relax}%
\providecommand \@@startlink[1]{}%
\providecommand \@@endlink[0]{}%
\providecommand \url  [0]{\begingroup\@sanitize@url \@url }%
\providecommand \@url [1]{\endgroup\@href {#1}{\urlprefix }}%
\providecommand \urlprefix  [0]{URL }%
\providecommand \Eprint [0]{\href }%
\providecommand \doibase [0]{http://dx.doi.org/}%
\providecommand \selectlanguage [0]{\@gobble}%
\providecommand \bibinfo  [0]{\@secondoftwo}%
\providecommand \bibfield  [0]{\@secondoftwo}%
\providecommand \translation [1]{[#1]}%
\providecommand \BibitemOpen [0]{}%
\providecommand \bibitemStop [0]{}%
\providecommand \bibitemNoStop [0]{.\EOS\space}%
\providecommand \EOS [0]{\spacefactor3000\relax}%
\providecommand \BibitemShut  [1]{\csname bibitem#1\endcsname}%
\let\auto@bib@innerbib\@empty
\bibitem [{\citenamefont {Zyla}\ \emph {et~al.}(2020)\citenamefont {Zyla} \emph
  {et~al.}}]{Zyla:2020zbs}%
  \BibitemOpen
  \bibfield  {author} {\bibinfo {author} {\bibfnamefont {P.}~\bibnamefont
  {Zyla}} \emph {et~al.} (\bibinfo {collaboration} {Particle Data Group}),\
  }\href {\doibase 10.1093/ptep/ptaa104} {\bibfield  {journal} {\bibinfo
  {journal} {PTEP}\ }\textbf {\bibinfo {volume} {2020}},\ \bibinfo {pages}
  {083C01} (\bibinfo {year} {2020})}\BibitemShut {NoStop}%
\bibitem [{\citenamefont {Duan}\ \emph
  {et~al.}(2006{\natexlab{a}})\citenamefont {Duan}, \citenamefont {Fuller},\
  and\ \citenamefont {Qian}}]{Duan:2005cp}%
  \BibitemOpen
  \bibfield  {author} {\bibinfo {author} {\bibfnamefont {H.}~\bibnamefont
  {Duan}}, \bibinfo {author} {\bibfnamefont {G.~M.}\ \bibnamefont {Fuller}}, \
  and\ \bibinfo {author} {\bibfnamefont {Y.-Z.}\ \bibnamefont {Qian}},\ }\href
  {\doibase 10.1103/PhysRevD.74.123004} {\bibfield  {journal} {\bibinfo
  {journal} {Phys. Rev. D}\ }\textbf {\bibinfo {volume} {74}},\ \bibinfo
  {pages} {123004} (\bibinfo {year} {2006}{\natexlab{a}})},\ \Eprint
  {http://arxiv.org/abs/astro-ph/0511275} {arXiv:astro-ph/0511275} \BibitemShut
  {NoStop}%
\bibitem [{\citenamefont {Duan}\ \emph
  {et~al.}(2006{\natexlab{b}})\citenamefont {Duan}, \citenamefont {Fuller},
  \citenamefont {Carlson},\ and\ \citenamefont {Qian}}]{Duan:2006an}%
  \BibitemOpen
  \bibfield  {author} {\bibinfo {author} {\bibfnamefont {H.}~\bibnamefont
  {Duan}}, \bibinfo {author} {\bibfnamefont {G.~M.}\ \bibnamefont {Fuller}},
  \bibinfo {author} {\bibfnamefont {J.}~\bibnamefont {Carlson}}, \ and\
  \bibinfo {author} {\bibfnamefont {Y.-Z.}\ \bibnamefont {Qian}},\ }\href
  {\doibase 10.1103/PhysRevD.74.105014} {\bibfield  {journal} {\bibinfo
  {journal} {Phys. Rev.}\ }\textbf {\bibinfo {volume} {D74}},\ \bibinfo {pages}
  {105014} (\bibinfo {year} {2006}{\natexlab{b}})},\ \Eprint
  {http://arxiv.org/abs/astro-ph/0606616} {arXiv:astro-ph/0606616 [astro-ph]}
  \BibitemShut {NoStop}%
\bibitem [{\citenamefont {Hannestad}\ \emph {et~al.}(2006)\citenamefont
  {Hannestad}, \citenamefont {Raffelt}, \citenamefont {Sigl},\ and\
  \citenamefont {Wong}}]{Hannestad:2006nj}%
  \BibitemOpen
  \bibfield  {author} {\bibinfo {author} {\bibfnamefont {S.}~\bibnamefont
  {Hannestad}}, \bibinfo {author} {\bibfnamefont {G.~G.}\ \bibnamefont
  {Raffelt}}, \bibinfo {author} {\bibfnamefont {G.}~\bibnamefont {Sigl}}, \
  and\ \bibinfo {author} {\bibfnamefont {Y.~Y.~Y.}\ \bibnamefont {Wong}},\
  }\href {\doibase 10.1103/PhysRevD.74.105010} {\bibfield  {journal} {\bibinfo
  {journal} {Phys. Rev. D}\ }\textbf {\bibinfo {volume} {74}},\ \bibinfo
  {pages} {105010} (\bibinfo {year} {2006})},\ \bibinfo {note} {[Erratum:
  Phys.Rev.D 76, 029901 (2007)]},\ \Eprint
  {http://arxiv.org/abs/astro-ph/0608695} {arXiv:astro-ph/0608695} \BibitemShut
  {NoStop}%
\bibitem [{\citenamefont {Esteban-Pretel}\ \emph {et~al.}(2008)\citenamefont
  {Esteban-Pretel}, \citenamefont {Mirizzi}, \citenamefont {Pastor},
  \citenamefont {Tomas}, \citenamefont {Raffelt}, \citenamefont {Serpico},\
  and\ \citenamefont {Sigl}}]{Esteban-Pretel:2008ovd}%
  \BibitemOpen
  \bibfield  {author} {\bibinfo {author} {\bibfnamefont {A.}~\bibnamefont
  {Esteban-Pretel}}, \bibinfo {author} {\bibfnamefont {A.}~\bibnamefont
  {Mirizzi}}, \bibinfo {author} {\bibfnamefont {S.}~\bibnamefont {Pastor}},
  \bibinfo {author} {\bibfnamefont {R.}~\bibnamefont {Tomas}}, \bibinfo
  {author} {\bibfnamefont {G.~G.}\ \bibnamefont {Raffelt}}, \bibinfo {author}
  {\bibfnamefont {P.~D.}\ \bibnamefont {Serpico}}, \ and\ \bibinfo {author}
  {\bibfnamefont {G.}~\bibnamefont {Sigl}},\ }\href {\doibase
  10.1103/PhysRevD.78.085012} {\bibfield  {journal} {\bibinfo  {journal} {Phys.
  Rev. D}\ }\textbf {\bibinfo {volume} {78}},\ \bibinfo {pages} {085012}
  (\bibinfo {year} {2008})},\ \Eprint {http://arxiv.org/abs/0807.0659}
  {arXiv:0807.0659 [astro-ph]} \BibitemShut {NoStop}%
\bibitem [{\citenamefont {Dasgupta}\ \emph {et~al.}(2009)\citenamefont
  {Dasgupta}, \citenamefont {Dighe}, \citenamefont {Raffelt},\ and\
  \citenamefont {Smirnov}}]{Dasgupta:2009mg}%
  \BibitemOpen
  \bibfield  {author} {\bibinfo {author} {\bibfnamefont {B.}~\bibnamefont
  {Dasgupta}}, \bibinfo {author} {\bibfnamefont {A.}~\bibnamefont {Dighe}},
  \bibinfo {author} {\bibfnamefont {G.~G.}\ \bibnamefont {Raffelt}}, \ and\
  \bibinfo {author} {\bibfnamefont {A.~Y.}\ \bibnamefont {Smirnov}},\ }\href
  {\doibase 10.1103/PhysRevLett.103.051105} {\bibfield  {journal} {\bibinfo
  {journal} {Phys. Rev. Lett.}\ }\textbf {\bibinfo {volume} {103}},\ \bibinfo
  {pages} {051105} (\bibinfo {year} {2009})},\ \Eprint
  {http://arxiv.org/abs/0904.3542} {arXiv:0904.3542 [hep-ph]} \BibitemShut
  {NoStop}%
\bibitem [{\citenamefont {Malkus}\ \emph {et~al.}(2012)\citenamefont {Malkus},
  \citenamefont {Kneller}, \citenamefont {McLaughlin},\ and\ \citenamefont
  {Surman}}]{Malkus:2012ts}%
  \BibitemOpen
  \bibfield  {author} {\bibinfo {author} {\bibfnamefont {A.}~\bibnamefont
  {Malkus}}, \bibinfo {author} {\bibfnamefont {J.~P.}\ \bibnamefont {Kneller}},
  \bibinfo {author} {\bibfnamefont {G.~C.}\ \bibnamefont {McLaughlin}}, \ and\
  \bibinfo {author} {\bibfnamefont {R.}~\bibnamefont {Surman}},\ }\href
  {\doibase 10.1103/PhysRevD.86.085015} {\bibfield  {journal} {\bibinfo
  {journal} {Phys. Rev. D}\ }\textbf {\bibinfo {volume} {86}},\ \bibinfo
  {pages} {085015} (\bibinfo {year} {2012})},\ \Eprint
  {http://arxiv.org/abs/1207.6648} {arXiv:1207.6648 [hep-ph]} \BibitemShut
  {NoStop}%
\bibitem [{\citenamefont {Cherry}\ \emph {et~al.}(2012)\citenamefont {Cherry},
  \citenamefont {Carlson}, \citenamefont {Friedland}, \citenamefont {Fuller},\
  and\ \citenamefont {Vlasenko}}]{Cherry:2012zw}%
  \BibitemOpen
  \bibfield  {author} {\bibinfo {author} {\bibfnamefont {J.~F.}\ \bibnamefont
  {Cherry}}, \bibinfo {author} {\bibfnamefont {J.}~\bibnamefont {Carlson}},
  \bibinfo {author} {\bibfnamefont {A.}~\bibnamefont {Friedland}}, \bibinfo
  {author} {\bibfnamefont {G.~M.}\ \bibnamefont {Fuller}}, \ and\ \bibinfo
  {author} {\bibfnamefont {A.}~\bibnamefont {Vlasenko}},\ }\href {\doibase
  10.1103/PhysRevLett.108.261104} {\bibfield  {journal} {\bibinfo  {journal}
  {Phys. Rev. Lett.}\ }\textbf {\bibinfo {volume} {108}},\ \bibinfo {pages}
  {261104} (\bibinfo {year} {2012})},\ \Eprint {http://arxiv.org/abs/1203.1607}
  {arXiv:1203.1607 [hep-ph]} \BibitemShut {NoStop}%
\bibitem [{\citenamefont {Raffelt}\ \emph {et~al.}(2013)\citenamefont
  {Raffelt}, \citenamefont {Sarikas},\ and\ \citenamefont
  {de~Sousa~Seixas}}]{Raffelt:2013rqa}%
  \BibitemOpen
  \bibfield  {author} {\bibinfo {author} {\bibfnamefont {G.}~\bibnamefont
  {Raffelt}}, \bibinfo {author} {\bibfnamefont {S.}~\bibnamefont {Sarikas}}, \
  and\ \bibinfo {author} {\bibfnamefont {D.}~\bibnamefont {de~Sousa~Seixas}},\
  }\href {\doibase 10.1103/PhysRevLett.113.239903,
  10.1103/PhysRevLett.111.091101} {\bibfield  {journal} {\bibinfo  {journal}
  {Phys. Rev. Lett.}\ }\textbf {\bibinfo {volume} {111}},\ \bibinfo {pages}
  {091101} (\bibinfo {year} {2013})},\ \bibinfo {note} {[Erratum: Phys. Rev.
  Lett.113,no.23,239903(2014)]},\ \Eprint {http://arxiv.org/abs/1305.7140}
  {arXiv:1305.7140 [hep-ph]} \BibitemShut {NoStop}%
\bibitem [{\citenamefont {Vlasenko}\ \emph {et~al.}(2014)\citenamefont
  {Vlasenko}, \citenamefont {Fuller},\ and\ \citenamefont
  {Cirigliano}}]{Vlasenko:2013fja}%
  \BibitemOpen
  \bibfield  {author} {\bibinfo {author} {\bibfnamefont {A.}~\bibnamefont
  {Vlasenko}}, \bibinfo {author} {\bibfnamefont {G.~M.}\ \bibnamefont
  {Fuller}}, \ and\ \bibinfo {author} {\bibfnamefont {V.}~\bibnamefont
  {Cirigliano}},\ }\href {\doibase 10.1103/PhysRevD.89.105004} {\bibfield
  {journal} {\bibinfo  {journal} {Phys. Rev.}\ }\textbf {\bibinfo {volume}
  {D89}},\ \bibinfo {pages} {105004} (\bibinfo {year} {2014})}\BibitemShut
  {NoStop}%
\bibitem [{\citenamefont {Volpe}\ \emph {et~al.}(2013)\citenamefont {Volpe},
  \citenamefont {V{\"a\"an\"a}nen},\ and\ \citenamefont
  {Espinoza}}]{Volpe:2013uxl}%
  \BibitemOpen
  \bibfield  {author} {\bibinfo {author} {\bibfnamefont {C.}~\bibnamefont
  {Volpe}}, \bibinfo {author} {\bibfnamefont {D.}~\bibnamefont
  {V{\"a\"an\"a}nen}}, \ and\ \bibinfo {author} {\bibfnamefont
  {C.}~\bibnamefont {Espinoza}},\ }\href {\doibase 10.1103/PhysRevD.87.113010}
  {\bibfield  {journal} {\bibinfo  {journal} {Phys.Rev.}\ }\textbf {\bibinfo
  {volume} {D87}},\ \bibinfo {pages} {113010} (\bibinfo {year}
  {2013})}\BibitemShut {NoStop}%
\bibitem [{\citenamefont {Wu}\ \emph {et~al.}(2016)\citenamefont {Wu},
  \citenamefont {Duan},\ and\ \citenamefont {Qian}}]{Wu:2015fga}%
  \BibitemOpen
  \bibfield  {author} {\bibinfo {author} {\bibfnamefont {M.-R.}\ \bibnamefont
  {Wu}}, \bibinfo {author} {\bibfnamefont {H.}~\bibnamefont {Duan}}, \ and\
  \bibinfo {author} {\bibfnamefont {Y.-Z.}\ \bibnamefont {Qian}},\ }\href
  {\doibase 10.1016/j.physletb.2015.11.027} {\bibfield  {journal} {\bibinfo
  {journal} {Phys. Lett.}\ }\textbf {\bibinfo {volume} {B752}},\ \bibinfo
  {pages} {89} (\bibinfo {year} {2016})},\ \Eprint
  {http://arxiv.org/abs/1509.08975} {arXiv:1509.08975 [hep-ph]} \BibitemShut
  {NoStop}%
\bibitem [{\citenamefont {Abbar}\ and\ \citenamefont
  {Duan}(2015)}]{Abbar:2015fwa}%
  \BibitemOpen
  \bibfield  {author} {\bibinfo {author} {\bibfnamefont {S.}~\bibnamefont
  {Abbar}}\ and\ \bibinfo {author} {\bibfnamefont {H.}~\bibnamefont {Duan}},\
  }\href {\doibase 10.1016/j.physletb.2015.10.019} {\bibfield  {journal}
  {\bibinfo  {journal} {Phys. Lett. B}\ }\textbf {\bibinfo {volume} {751}},\
  \bibinfo {pages} {43} (\bibinfo {year} {2015})},\ \Eprint
  {http://arxiv.org/abs/1509.01538} {arXiv:1509.01538 [astro-ph.HE]}
  \BibitemShut {NoStop}%
\bibitem [{\citenamefont {Sawyer}(2016)}]{Sawyer:2015dsa}%
  \BibitemOpen
  \bibfield  {author} {\bibinfo {author} {\bibfnamefont {R.~F.}\ \bibnamefont
  {Sawyer}},\ }\href {\doibase 10.1103/PhysRevLett.116.081101} {\bibfield
  {journal} {\bibinfo  {journal} {Phys. Rev. Lett.}\ }\textbf {\bibinfo
  {volume} {116}},\ \bibinfo {pages} {081101} (\bibinfo {year} {2016})},\
  \Eprint {http://arxiv.org/abs/1509.03323} {arXiv:1509.03323 [astro-ph.HE]}
  \BibitemShut {NoStop}%
\bibitem [{\citenamefont {Izaguirre}\ \emph {et~al.}(2017)\citenamefont
  {Izaguirre}, \citenamefont {Raffelt},\ and\ \citenamefont
  {Tamborra}}]{Izaguirre:2016gsx}%
  \BibitemOpen
  \bibfield  {author} {\bibinfo {author} {\bibfnamefont {I.}~\bibnamefont
  {Izaguirre}}, \bibinfo {author} {\bibfnamefont {G.}~\bibnamefont {Raffelt}},
  \ and\ \bibinfo {author} {\bibfnamefont {I.}~\bibnamefont {Tamborra}},\
  }\href {\doibase 10.1103/PhysRevLett.118.021101} {\bibfield  {journal}
  {\bibinfo  {journal} {Phys. Rev. Lett.}\ }\textbf {\bibinfo {volume} {118}},\
  \bibinfo {pages} {021101} (\bibinfo {year} {2017})},\ \Eprint
  {http://arxiv.org/abs/1610.01612} {arXiv:1610.01612 [hep-ph]} \BibitemShut
  {NoStop}%
\bibitem [{\citenamefont {Capozzi}\ \emph {et~al.}(2019)\citenamefont
  {Capozzi}, \citenamefont {Dasgupta}, \citenamefont {Mirizzi}, \citenamefont
  {Sen},\ and\ \citenamefont {Sigl}}]{Capozzi:2018clo}%
  \BibitemOpen
  \bibfield  {author} {\bibinfo {author} {\bibfnamefont {F.}~\bibnamefont
  {Capozzi}}, \bibinfo {author} {\bibfnamefont {B.}~\bibnamefont {Dasgupta}},
  \bibinfo {author} {\bibfnamefont {A.}~\bibnamefont {Mirizzi}}, \bibinfo
  {author} {\bibfnamefont {M.}~\bibnamefont {Sen}}, \ and\ \bibinfo {author}
  {\bibfnamefont {G.}~\bibnamefont {Sigl}},\ }\href {\doibase
  10.1103/PhysRevLett.122.091101} {\bibfield  {journal} {\bibinfo  {journal}
  {Phys. Rev. Lett.}\ }\textbf {\bibinfo {volume} {122}},\ \bibinfo {pages}
  {091101} (\bibinfo {year} {2019})},\ \Eprint
  {http://arxiv.org/abs/1808.06618} {arXiv:1808.06618 [hep-ph]} \BibitemShut
  {NoStop}%
\bibitem [{\citenamefont {Abbar}(2021)}]{Abbar:2020ror}%
  \BibitemOpen
  \bibfield  {author} {\bibinfo {author} {\bibfnamefont {S.}~\bibnamefont
  {Abbar}},\ }\href {\doibase 10.1103/PhysRevD.103.045014} {\bibfield
  {journal} {\bibinfo  {journal} {Phys. Rev. D}\ }\textbf {\bibinfo {volume}
  {103}},\ \bibinfo {pages} {045014} (\bibinfo {year} {2021})},\ \Eprint
  {http://arxiv.org/abs/2007.13655} {arXiv:2007.13655 [astro-ph.HE]}
  \BibitemShut {NoStop}%
\bibitem [{\citenamefont {Xiong}\ and\ \citenamefont
  {Qian}(2021)}]{Xiong:2021dex}%
  \BibitemOpen
  \bibfield  {author} {\bibinfo {author} {\bibfnamefont {Z.}~\bibnamefont
  {Xiong}}\ and\ \bibinfo {author} {\bibfnamefont {Y.-Z.}\ \bibnamefont
  {Qian}},\ }\href {\doibase 10.1016/j.physletb.2021.136550} {\bibfield
  {journal} {\bibinfo  {journal} {Phys. Lett. B}\ }\textbf {\bibinfo {volume}
  {820}},\ \bibinfo {pages} {136550} (\bibinfo {year} {2021})},\ \Eprint
  {http://arxiv.org/abs/2104.05618} {arXiv:2104.05618 [astro-ph.HE]}
  \BibitemShut {NoStop}%
\bibitem [{\citenamefont {Johns}(2021)}]{Johns:2021qby}%
  \BibitemOpen
  \bibfield  {author} {\bibinfo {author} {\bibfnamefont {L.}~\bibnamefont
  {Johns}},\ }\href@noop {} {\  (\bibinfo {year} {2021})},\ \Eprint
  {http://arxiv.org/abs/2104.11369} {arXiv:2104.11369 [hep-ph]} \BibitemShut
  {NoStop}%
\bibitem [{\citenamefont {Duan}\ \emph {et~al.}(2010)\citenamefont {Duan},
  \citenamefont {Fuller},\ and\ \citenamefont {Qian}}]{Duan:2010bg}%
  \BibitemOpen
  \bibfield  {author} {\bibinfo {author} {\bibfnamefont {H.}~\bibnamefont
  {Duan}}, \bibinfo {author} {\bibfnamefont {G.~M.}\ \bibnamefont {Fuller}}, \
  and\ \bibinfo {author} {\bibfnamefont {Y.-Z.}\ \bibnamefont {Qian}},\ }\href
  {\doibase 10.1146/annurev.nucl.012809.104524} {\bibfield  {journal} {\bibinfo
   {journal} {Ann. Rev. Nucl. Part. Sci.}\ }\textbf {\bibinfo {volume} {60}},\
  \bibinfo {pages} {569} (\bibinfo {year} {2010})}\BibitemShut {NoStop}%
\bibitem [{\citenamefont {Mirizzi}\ \emph {et~al.}(2015)\citenamefont
  {Mirizzi}, \citenamefont {Mangano},\ and\ \citenamefont
  {Saviano}}]{Mirizzi:2015fva}%
  \BibitemOpen
  \bibfield  {author} {\bibinfo {author} {\bibfnamefont {A.}~\bibnamefont
  {Mirizzi}}, \bibinfo {author} {\bibfnamefont {G.}~\bibnamefont {Mangano}}, \
  and\ \bibinfo {author} {\bibfnamefont {N.}~\bibnamefont {Saviano}},\ }\href
  {\doibase 10.1103/PhysRevD.92.021702} {\bibfield  {journal} {\bibinfo
  {journal} {Phys. Rev.}\ }\textbf {\bibinfo {volume} {D92}},\ \bibinfo {pages}
  {021702} (\bibinfo {year} {2015})},\ \Eprint
  {http://arxiv.org/abs/1503.03485} {arXiv:1503.03485 [hep-ph]} \BibitemShut
  {NoStop}%
\bibitem [{\citenamefont {Duan}(2015)}]{Duan:2015cqa}%
  \BibitemOpen
  \bibfield  {author} {\bibinfo {author} {\bibfnamefont {H.}~\bibnamefont
  {Duan}},\ }\href {\doibase 10.1142/S0218301315410086} {\bibfield  {journal}
  {\bibinfo  {journal} {Int. J. Mod. Phys.}\ }\textbf {\bibinfo {volume}
  {E24}},\ \bibinfo {pages} {1541008} (\bibinfo {year} {2015})},\ \Eprint
  {http://arxiv.org/abs/1506.08629} {arXiv:1506.08629 [hep-ph]} \BibitemShut
  {NoStop}%
\bibitem [{\citenamefont {Tamborra}\ and\ \citenamefont
  {Shalgar}(2020)}]{Tamborra:2020cul}%
  \BibitemOpen
  \bibfield  {author} {\bibinfo {author} {\bibfnamefont {I.}~\bibnamefont
  {Tamborra}}\ and\ \bibinfo {author} {\bibfnamefont {S.}~\bibnamefont
  {Shalgar}},\ }\href {\doibase 10.1146/annurev-nucl-102920-050505} {\
  (\bibinfo {year} {2020}),\ 10.1146/annurev-nucl-102920-050505},\ \Eprint
  {http://arxiv.org/abs/2011.01948} {arXiv:2011.01948 [astro-ph.HE]}
  \BibitemShut {NoStop}%
\bibitem [{\citenamefont {Duan}\ \emph {et~al.}(2011)\citenamefont {Duan},
  \citenamefont {Friedland}, \citenamefont {McLaughlin},\ and\ \citenamefont
  {Surman}}]{Duan:2010af}%
  \BibitemOpen
  \bibfield  {author} {\bibinfo {author} {\bibfnamefont {H.}~\bibnamefont
  {Duan}}, \bibinfo {author} {\bibfnamefont {A.}~\bibnamefont {Friedland}},
  \bibinfo {author} {\bibfnamefont {G.~C.}\ \bibnamefont {McLaughlin}}, \ and\
  \bibinfo {author} {\bibfnamefont {R.}~\bibnamefont {Surman}},\ }\href
  {\doibase 10.1088/0954-3899/38/3/035201} {\bibfield  {journal} {\bibinfo
  {journal} {J. Phys.}\ }\textbf {\bibinfo {volume} {G38}},\ \bibinfo {pages}
  {035201} (\bibinfo {year} {2011})}\BibitemShut {NoStop}%
\bibitem [{\citenamefont {Wu}\ \emph {et~al.}(2015)\citenamefont {Wu},
  \citenamefont {Qian}, \citenamefont {Martinez-Pinedo}, \citenamefont
  {Fischer},\ and\ \citenamefont {Huther}}]{Wu:2014kaa}%
  \BibitemOpen
  \bibfield  {author} {\bibinfo {author} {\bibfnamefont {M.-R.}\ \bibnamefont
  {Wu}}, \bibinfo {author} {\bibfnamefont {Y.-Z.}\ \bibnamefont {Qian}},
  \bibinfo {author} {\bibfnamefont {G.}~\bibnamefont {Martinez-Pinedo}},
  \bibinfo {author} {\bibfnamefont {T.}~\bibnamefont {Fischer}}, \ and\
  \bibinfo {author} {\bibfnamefont {L.}~\bibnamefont {Huther}},\ }\href@noop {}
  {\bibfield  {journal} {\bibinfo  {journal} {Phys. Rev.}\ }\textbf {\bibinfo
  {volume} {D91}},\ \bibinfo {pages} {065016} (\bibinfo {year}
  {2015})}\BibitemShut {NoStop}%
\bibitem [{\citenamefont {Sasaki}\ \emph {et~al.}(2017)\citenamefont {Sasaki},
  \citenamefont {Kajino}, \citenamefont {Takiwaki}, \citenamefont {Hayakawa},
  \citenamefont {Balantekin},\ and\ \citenamefont {Pehlivan}}]{Sasaki:2017jry}%
  \BibitemOpen
  \bibfield  {author} {\bibinfo {author} {\bibfnamefont {H.}~\bibnamefont
  {Sasaki}}, \bibinfo {author} {\bibfnamefont {T.}~\bibnamefont {Kajino}},
  \bibinfo {author} {\bibfnamefont {T.}~\bibnamefont {Takiwaki}}, \bibinfo
  {author} {\bibfnamefont {T.}~\bibnamefont {Hayakawa}}, \bibinfo {author}
  {\bibfnamefont {A.~B.}\ \bibnamefont {Balantekin}}, \ and\ \bibinfo {author}
  {\bibfnamefont {Y.}~\bibnamefont {Pehlivan}},\ }\href {\doibase
  10.1103/PhysRevD.96.043013} {\bibfield  {journal} {\bibinfo  {journal} {Phys.
  Rev. D}\ }\textbf {\bibinfo {volume} {96}},\ \bibinfo {pages} {043013}
  (\bibinfo {year} {2017})},\ \Eprint {http://arxiv.org/abs/1707.09111}
  {arXiv:1707.09111 [astro-ph.HE]} \BibitemShut {NoStop}%
\bibitem [{\citenamefont {Wu}\ \emph {et~al.}(2017)\citenamefont {Wu},
  \citenamefont {Tamborra}, \citenamefont {Just},\ and\ \citenamefont
  {Janka}}]{Wu:2017drk}%
  \BibitemOpen
  \bibfield  {author} {\bibinfo {author} {\bibfnamefont {M.-R.}\ \bibnamefont
  {Wu}}, \bibinfo {author} {\bibfnamefont {I.}~\bibnamefont {Tamborra}},
  \bibinfo {author} {\bibfnamefont {O.}~\bibnamefont {Just}}, \ and\ \bibinfo
  {author} {\bibfnamefont {H.-T.}\ \bibnamefont {Janka}},\ }\href {\doibase
  10.1103/PhysRevD.96.123015} {\bibfield  {journal} {\bibinfo  {journal} {Phys.
  Rev.}\ }\textbf {\bibinfo {volume} {D96}},\ \bibinfo {pages} {123015}
  (\bibinfo {year} {2017})},\ \Eprint {http://arxiv.org/abs/1711.00477}
  {arXiv:1711.00477 [astro-ph.HE]} \BibitemShut {NoStop}%
\bibitem [{\citenamefont {Wu}\ and\ \citenamefont
  {Tamborra}(2017)}]{Wu:2017qpc}%
  \BibitemOpen
  \bibfield  {author} {\bibinfo {author} {\bibfnamefont {M.-R.}\ \bibnamefont
  {Wu}}\ and\ \bibinfo {author} {\bibfnamefont {I.}~\bibnamefont {Tamborra}},\
  }\href {\doibase 10.1103/PhysRevD.95.103007} {\bibfield  {journal} {\bibinfo
  {journal} {Phys. Rev.}\ }\textbf {\bibinfo {volume} {D95}},\ \bibinfo {pages}
  {103007} (\bibinfo {year} {2017})},\ \Eprint
  {http://arxiv.org/abs/1701.06580} {arXiv:1701.06580 [astro-ph.HE]}
  \BibitemShut {NoStop}%
\bibitem [{\citenamefont {Stapleford}\ \emph {et~al.}(2020)\citenamefont
  {Stapleford}, \citenamefont {Fr\"ohlich},\ and\ \citenamefont
  {Kneller}}]{Stapleford:2019yqg}%
  \BibitemOpen
  \bibfield  {author} {\bibinfo {author} {\bibfnamefont {C.~J.}\ \bibnamefont
  {Stapleford}}, \bibinfo {author} {\bibfnamefont {C.}~\bibnamefont
  {Fr\"ohlich}}, \ and\ \bibinfo {author} {\bibfnamefont {J.~P.}\ \bibnamefont
  {Kneller}},\ }\href {\doibase 10.1103/PhysRevD.102.081301} {\bibfield
  {journal} {\bibinfo  {journal} {Phys. Rev. D}\ }\textbf {\bibinfo {volume}
  {102}},\ \bibinfo {pages} {081301} (\bibinfo {year} {2020})},\ \Eprint
  {http://arxiv.org/abs/1910.04172} {arXiv:1910.04172 [astro-ph.HE]}
  \BibitemShut {NoStop}%
\bibitem [{\citenamefont {Xiong}\ \emph {et~al.}(2020)\citenamefont {Xiong},
  \citenamefont {Sieverding}, \citenamefont {Sen},\ and\ \citenamefont
  {Qian}}]{Xiong:2020ntn}%
  \BibitemOpen
  \bibfield  {author} {\bibinfo {author} {\bibfnamefont {Z.}~\bibnamefont
  {Xiong}}, \bibinfo {author} {\bibfnamefont {A.}~\bibnamefont {Sieverding}},
  \bibinfo {author} {\bibfnamefont {M.}~\bibnamefont {Sen}}, \ and\ \bibinfo
  {author} {\bibfnamefont {Y.-Z.}\ \bibnamefont {Qian}},\ }\href {\doibase
  10.3847/1538-4357/abac5e} {\bibfield  {journal} {\bibinfo  {journal}
  {Astrophys. J.}\ }\textbf {\bibinfo {volume} {900}},\ \bibinfo {pages} {144}
  (\bibinfo {year} {2020})},\ \Eprint {http://arxiv.org/abs/2006.11414}
  {arXiv:2006.11414 [astro-ph.HE]} \BibitemShut {NoStop}%
\bibitem [{\citenamefont {George}\ \emph {et~al.}(2020)\citenamefont {George},
  \citenamefont {Wu}, \citenamefont {Tamborra}, \citenamefont
  {Ardevol-Pulpillo},\ and\ \citenamefont {Janka}}]{George:2020veu}%
  \BibitemOpen
  \bibfield  {author} {\bibinfo {author} {\bibfnamefont {M.}~\bibnamefont
  {George}}, \bibinfo {author} {\bibfnamefont {M.-R.}\ \bibnamefont {Wu}},
  \bibinfo {author} {\bibfnamefont {I.}~\bibnamefont {Tamborra}}, \bibinfo
  {author} {\bibfnamefont {R.}~\bibnamefont {Ardevol-Pulpillo}}, \ and\
  \bibinfo {author} {\bibfnamefont {H.-T.}\ \bibnamefont {Janka}},\ }\href
  {\doibase 10.1103/PhysRevD.102.103015} {\bibfield  {journal} {\bibinfo
  {journal} {Phys. Rev. D}\ }\textbf {\bibinfo {volume} {102}},\ \bibinfo
  {pages} {103015} (\bibinfo {year} {2020})},\ \Eprint
  {http://arxiv.org/abs/2009.04046} {arXiv:2009.04046 [astro-ph.HE]}
  \BibitemShut {NoStop}%
\bibitem [{\citenamefont {Li}\ and\ \citenamefont {Siegel}(2021)}]{Li:2021vqj}%
  \BibitemOpen
  \bibfield  {author} {\bibinfo {author} {\bibfnamefont {X.}~\bibnamefont
  {Li}}\ and\ \bibinfo {author} {\bibfnamefont {D.~M.}\ \bibnamefont
  {Siegel}},\ }\href {\doibase 10.1103/PhysRevLett.126.251101} {\bibfield
  {journal} {\bibinfo  {journal} {Phys. Rev. Lett.}\ }\textbf {\bibinfo
  {volume} {126}},\ \bibinfo {pages} {251101} (\bibinfo {year} {2021})},\
  \Eprint {http://arxiv.org/abs/2103.02616} {arXiv:2103.02616 [astro-ph.HE]}
  \BibitemShut {NoStop}%
\bibitem [{\citenamefont {Dasgupta}\ \emph {et~al.}(2017)\citenamefont
  {Dasgupta}, \citenamefont {Mirizzi},\ and\ \citenamefont
  {Sen}}]{Dasgupta:2016dbv}%
  \BibitemOpen
  \bibfield  {author} {\bibinfo {author} {\bibfnamefont {B.}~\bibnamefont
  {Dasgupta}}, \bibinfo {author} {\bibfnamefont {A.}~\bibnamefont {Mirizzi}}, \
  and\ \bibinfo {author} {\bibfnamefont {M.}~\bibnamefont {Sen}},\ }\href
  {\doibase 10.1088/1475-7516/2017/02/019} {\bibfield  {journal} {\bibinfo
  {journal} {JCAP}\ }\textbf {\bibinfo {volume} {1702}},\ \bibinfo {pages}
  {019} (\bibinfo {year} {2017})},\ \Eprint {http://arxiv.org/abs/1609.00528}
  {arXiv:1609.00528 [hep-ph]} \BibitemShut {NoStop}%
\bibitem [{\citenamefont {Capozzi}\ \emph {et~al.}(2017)\citenamefont
  {Capozzi}, \citenamefont {Dasgupta}, \citenamefont {Lisi}, \citenamefont
  {Marrone},\ and\ \citenamefont {Mirizzi}}]{Capozzi:2017gqd}%
  \BibitemOpen
  \bibfield  {author} {\bibinfo {author} {\bibfnamefont {F.}~\bibnamefont
  {Capozzi}}, \bibinfo {author} {\bibfnamefont {B.}~\bibnamefont {Dasgupta}},
  \bibinfo {author} {\bibfnamefont {E.}~\bibnamefont {Lisi}}, \bibinfo {author}
  {\bibfnamefont {A.}~\bibnamefont {Marrone}}, \ and\ \bibinfo {author}
  {\bibfnamefont {A.}~\bibnamefont {Mirizzi}},\ }\href {\doibase
  10.1103/PhysRevD.96.043016} {\bibfield  {journal} {\bibinfo  {journal} {Phys.
  Rev. D}\ }\textbf {\bibinfo {volume} {96}},\ \bibinfo {pages} {043016}
  (\bibinfo {year} {2017})},\ \Eprint {http://arxiv.org/abs/1706.03360}
  {arXiv:1706.03360 [hep-ph]} \BibitemShut {NoStop}%
\bibitem [{\citenamefont {Abbar}\ and\ \citenamefont
  {Volpe}(2019)}]{Abbar:2018beu}%
  \BibitemOpen
  \bibfield  {author} {\bibinfo {author} {\bibfnamefont {S.}~\bibnamefont
  {Abbar}}\ and\ \bibinfo {author} {\bibfnamefont {M.~C.}\ \bibnamefont
  {Volpe}},\ }\href {\doibase 10.1016/j.physletb.2019.02.002} {\bibfield
  {journal} {\bibinfo  {journal} {Phys. Lett. B}\ }\textbf {\bibinfo {volume}
  {790}},\ \bibinfo {pages} {545} (\bibinfo {year} {2019})},\ \Eprint
  {http://arxiv.org/abs/1811.04215} {arXiv:1811.04215 [astro-ph.HE]}
  \BibitemShut {NoStop}%
\bibitem [{\citenamefont {Yi}\ \emph {et~al.}(2019)\citenamefont {Yi},
  \citenamefont {Ma}, \citenamefont {Martin},\ and\ \citenamefont
  {Duan}}]{Yi:2019hrp}%
  \BibitemOpen
  \bibfield  {author} {\bibinfo {author} {\bibfnamefont {C.}~\bibnamefont
  {Yi}}, \bibinfo {author} {\bibfnamefont {L.}~\bibnamefont {Ma}}, \bibinfo
  {author} {\bibfnamefont {J.~D.}\ \bibnamefont {Martin}}, \ and\ \bibinfo
  {author} {\bibfnamefont {H.}~\bibnamefont {Duan}},\ }\href {\doibase
  10.1103/PhysRevD.99.063005} {\bibfield  {journal} {\bibinfo  {journal} {Phys.
  Rev. D}\ }\textbf {\bibinfo {volume} {99}},\ \bibinfo {pages} {063005}
  (\bibinfo {year} {2019})},\ \Eprint {http://arxiv.org/abs/1901.01546}
  {arXiv:1901.01546 [hep-ph]} \BibitemShut {NoStop}%
\bibitem [{\citenamefont {Martin}\ \emph {et~al.}(2020)\citenamefont {Martin},
  \citenamefont {Yi},\ and\ \citenamefont {Duan}}]{Martin:2019gxb}%
  \BibitemOpen
  \bibfield  {author} {\bibinfo {author} {\bibfnamefont {J.~D.}\ \bibnamefont
  {Martin}}, \bibinfo {author} {\bibfnamefont {C.}~\bibnamefont {Yi}}, \ and\
  \bibinfo {author} {\bibfnamefont {H.}~\bibnamefont {Duan}},\ }\href {\doibase
  10.1016/j.physletb.2019.135088} {\bibfield  {journal} {\bibinfo  {journal}
  {Phys. Lett. B}\ }\textbf {\bibinfo {volume} {800}},\ \bibinfo {pages}
  {135088} (\bibinfo {year} {2020})},\ \Eprint
  {http://arxiv.org/abs/1909.05225} {arXiv:1909.05225 [hep-ph]} \BibitemShut
  {NoStop}%
\bibitem [{\citenamefont {Padilla-Gay}\ \emph {et~al.}(2021)\citenamefont
  {Padilla-Gay}, \citenamefont {Shalgar},\ and\ \citenamefont
  {Tamborra}}]{Padilla-Gay:2020uxa}%
  \BibitemOpen
  \bibfield  {author} {\bibinfo {author} {\bibfnamefont {I.}~\bibnamefont
  {Padilla-Gay}}, \bibinfo {author} {\bibfnamefont {S.}~\bibnamefont
  {Shalgar}}, \ and\ \bibinfo {author} {\bibfnamefont {I.}~\bibnamefont
  {Tamborra}},\ }\href {\doibase 10.1088/1475-7516/2021/01/017} {\bibfield
  {journal} {\bibinfo  {journal} {JCAP}\ }\textbf {\bibinfo {volume} {01}},\
  \bibinfo {pages} {017} (\bibinfo {year} {2021})},\ \Eprint
  {http://arxiv.org/abs/2009.01843} {arXiv:2009.01843 [astro-ph.HE]}
  \BibitemShut {NoStop}%
\bibitem [{\citenamefont {Johns}\ \emph
  {et~al.}(2020{\natexlab{a}})\citenamefont {Johns}, \citenamefont {Nagakura},
  \citenamefont {Fuller},\ and\ \citenamefont {Burrows}}]{Johns:2020qsk}%
  \BibitemOpen
  \bibfield  {author} {\bibinfo {author} {\bibfnamefont {L.}~\bibnamefont
  {Johns}}, \bibinfo {author} {\bibfnamefont {H.}~\bibnamefont {Nagakura}},
  \bibinfo {author} {\bibfnamefont {G.~M.}\ \bibnamefont {Fuller}}, \ and\
  \bibinfo {author} {\bibfnamefont {A.}~\bibnamefont {Burrows}},\ }\href
  {\doibase 10.1103/PhysRevD.102.103017} {\bibfield  {journal} {\bibinfo
  {journal} {Phys. Rev. D}\ }\textbf {\bibinfo {volume} {102}},\ \bibinfo
  {pages} {103017} (\bibinfo {year} {2020}{\natexlab{a}})},\ \Eprint
  {http://arxiv.org/abs/2009.09024} {arXiv:2009.09024 [hep-ph]} \BibitemShut
  {NoStop}%
\bibitem [{\citenamefont {Bhattacharyya}\ and\ \citenamefont
  {Dasgupta}(2020)}]{Bhattacharyya:2020dhu}%
  \BibitemOpen
  \bibfield  {author} {\bibinfo {author} {\bibfnamefont {S.}~\bibnamefont
  {Bhattacharyya}}\ and\ \bibinfo {author} {\bibfnamefont {B.}~\bibnamefont
  {Dasgupta}},\ }\href {\doibase 10.1103/PhysRevD.102.063018} {\bibfield
  {journal} {\bibinfo  {journal} {Phys. Rev. D}\ }\textbf {\bibinfo {volume}
  {102}},\ \bibinfo {pages} {063018} (\bibinfo {year} {2020})},\ \Eprint
  {http://arxiv.org/abs/2005.00459} {arXiv:2005.00459 [hep-ph]} \BibitemShut
  {NoStop}%
\bibitem [{\citenamefont {Bhattacharyya}\ and\ \citenamefont
  {Dasgupta}(2021)}]{Bhattacharyya:2020jpj}%
  \BibitemOpen
  \bibfield  {author} {\bibinfo {author} {\bibfnamefont {S.}~\bibnamefont
  {Bhattacharyya}}\ and\ \bibinfo {author} {\bibfnamefont {B.}~\bibnamefont
  {Dasgupta}},\ }\href {\doibase 10.1103/PhysRevLett.126.061302} {\bibfield
  {journal} {\bibinfo  {journal} {Phys. Rev. Lett.}\ }\textbf {\bibinfo
  {volume} {126}},\ \bibinfo {pages} {061302} (\bibinfo {year} {2021})},\
  \Eprint {http://arxiv.org/abs/2009.03337} {arXiv:2009.03337 [hep-ph]}
  \BibitemShut {NoStop}%
\bibitem [{\citenamefont {Morinaga}(2021)}]{Morinaga:2021vmc}%
  \BibitemOpen
  \bibfield  {author} {\bibinfo {author} {\bibfnamefont {T.}~\bibnamefont
  {Morinaga}},\ }\href@noop {} {\  (\bibinfo {year} {2021})},\ \Eprint
  {http://arxiv.org/abs/2103.15267} {arXiv:2103.15267 [hep-ph]} \BibitemShut
  {NoStop}%
\bibitem [{\citenamefont {Richers}\ \emph {et~al.}(2021)\citenamefont
  {Richers}, \citenamefont {Willcox}, \citenamefont {Ford},\ and\ \citenamefont
  {Myers}}]{Richers:2021nbx}%
  \BibitemOpen
  \bibfield  {author} {\bibinfo {author} {\bibfnamefont {S.}~\bibnamefont
  {Richers}}, \bibinfo {author} {\bibfnamefont {D.~E.}\ \bibnamefont
  {Willcox}}, \bibinfo {author} {\bibfnamefont {N.~M.}\ \bibnamefont {Ford}}, \
  and\ \bibinfo {author} {\bibfnamefont {A.}~\bibnamefont {Myers}},\ }\href
  {\doibase 10.1103/PhysRevD.103.083013} {\bibfield  {journal} {\bibinfo
  {journal} {Phys. Rev. D}\ }\textbf {\bibinfo {volume} {103}},\ \bibinfo
  {pages} {083013} (\bibinfo {year} {2021})},\ \Eprint
  {http://arxiv.org/abs/2101.02745} {arXiv:2101.02745 [astro-ph.HE]}
  \BibitemShut {NoStop}%
\bibitem [{\citenamefont {Zaizen}\ and\ \citenamefont
  {Morinaga}(2021)}]{Zaizen:2021wwl}%
  \BibitemOpen
  \bibfield  {author} {\bibinfo {author} {\bibfnamefont {M.}~\bibnamefont
  {Zaizen}}\ and\ \bibinfo {author} {\bibfnamefont {T.}~\bibnamefont
  {Morinaga}},\ }\href@noop {} {\  (\bibinfo {year} {2021})},\ \Eprint
  {http://arxiv.org/abs/2104.10532} {arXiv:2104.10532 [hep-ph]} \BibitemShut
  {NoStop}%
\bibitem [{\citenamefont {Kato}\ \emph {et~al.}(2021)\citenamefont {Kato},
  \citenamefont {Nagakura},\ and\ \citenamefont {Morinaga}}]{Kato:2021cjf}%
  \BibitemOpen
  \bibfield  {author} {\bibinfo {author} {\bibfnamefont {C.}~\bibnamefont
  {Kato}}, \bibinfo {author} {\bibfnamefont {H.}~\bibnamefont {Nagakura}}, \
  and\ \bibinfo {author} {\bibfnamefont {T.}~\bibnamefont {Morinaga}},\
  }\href@noop {} {\  (\bibinfo {year} {2021})},\ \Eprint
  {http://arxiv.org/abs/2108.06356} {arXiv:2108.06356 [astro-ph.HE]}
  \BibitemShut {NoStop}%
\bibitem [{\citenamefont {Morinaga}\ \emph {et~al.}(2020)\citenamefont
  {Morinaga}, \citenamefont {Nagakura}, \citenamefont {Kato},\ and\
  \citenamefont {Yamada}}]{Morinaga:2019wsv}%
  \BibitemOpen
  \bibfield  {author} {\bibinfo {author} {\bibfnamefont {T.}~\bibnamefont
  {Morinaga}}, \bibinfo {author} {\bibfnamefont {H.}~\bibnamefont {Nagakura}},
  \bibinfo {author} {\bibfnamefont {C.}~\bibnamefont {Kato}}, \ and\ \bibinfo
  {author} {\bibfnamefont {S.}~\bibnamefont {Yamada}},\ }\href {\doibase
  10.1103/PhysRevResearch.2.012046} {\bibfield  {journal} {\bibinfo  {journal}
  {Phys. Rev. Res.}\ }\textbf {\bibinfo {volume} {2}},\ \bibinfo {pages}
  {012046} (\bibinfo {year} {2020})},\ \Eprint
  {http://arxiv.org/abs/1909.13131} {arXiv:1909.13131 [astro-ph.HE]}
  \BibitemShut {NoStop}%
\bibitem [{\citenamefont {Nagakura}\ \emph {et~al.}(2019)\citenamefont
  {Nagakura}, \citenamefont {Morinaga}, \citenamefont {Kato},\ and\
  \citenamefont {Yamada}}]{Nagakura:2019sig}%
  \BibitemOpen
  \bibfield  {author} {\bibinfo {author} {\bibfnamefont {H.}~\bibnamefont
  {Nagakura}}, \bibinfo {author} {\bibfnamefont {T.}~\bibnamefont {Morinaga}},
  \bibinfo {author} {\bibfnamefont {C.}~\bibnamefont {Kato}}, \ and\ \bibinfo
  {author} {\bibfnamefont {S.}~\bibnamefont {Yamada}},\ }\href {\doibase
  10.3847/1538-4357/ab4cf2} {\  (\bibinfo {year} {2019}),\
  10.3847/1538-4357/ab4cf2},\ \Eprint {http://arxiv.org/abs/1910.04288}
  {arXiv:1910.04288 [astro-ph.HE]} \BibitemShut {NoStop}%
\bibitem [{\citenamefont {Johns}\ \emph
  {et~al.}(2020{\natexlab{b}})\citenamefont {Johns}, \citenamefont {Nagakura},
  \citenamefont {Fuller},\ and\ \citenamefont {Burrows}}]{Johns:2019izj}%
  \BibitemOpen
  \bibfield  {author} {\bibinfo {author} {\bibfnamefont {L.}~\bibnamefont
  {Johns}}, \bibinfo {author} {\bibfnamefont {H.}~\bibnamefont {Nagakura}},
  \bibinfo {author} {\bibfnamefont {G.~M.}\ \bibnamefont {Fuller}}, \ and\
  \bibinfo {author} {\bibfnamefont {A.}~\bibnamefont {Burrows}},\ }\href
  {\doibase 10.1103/PhysRevD.101.043009} {\bibfield  {journal} {\bibinfo
  {journal} {Phys. Rev. D}\ }\textbf {\bibinfo {volume} {101}},\ \bibinfo
  {pages} {043009} (\bibinfo {year} {2020}{\natexlab{b}})},\ \Eprint
  {http://arxiv.org/abs/1910.05682} {arXiv:1910.05682 [hep-ph]} \BibitemShut
  {NoStop}%
\bibitem [{\citenamefont {Delfan~Azari}\ \emph {et~al.}(2020)\citenamefont
  {Delfan~Azari}, \citenamefont {Yamada}, \citenamefont {Morinaga},
  \citenamefont {Nagakura}, \citenamefont {Furusawa}, \citenamefont {Harada},
  \citenamefont {Okawa}, \citenamefont {Iwakami},\ and\ \citenamefont
  {Sumiyoshi}}]{DelfanAzari:2019tez}%
  \BibitemOpen
  \bibfield  {author} {\bibinfo {author} {\bibfnamefont {M.}~\bibnamefont
  {Delfan~Azari}}, \bibinfo {author} {\bibfnamefont {S.}~\bibnamefont
  {Yamada}}, \bibinfo {author} {\bibfnamefont {T.}~\bibnamefont {Morinaga}},
  \bibinfo {author} {\bibfnamefont {H.}~\bibnamefont {Nagakura}}, \bibinfo
  {author} {\bibfnamefont {S.}~\bibnamefont {Furusawa}}, \bibinfo {author}
  {\bibfnamefont {A.}~\bibnamefont {Harada}}, \bibinfo {author} {\bibfnamefont
  {H.}~\bibnamefont {Okawa}}, \bibinfo {author} {\bibfnamefont
  {W.}~\bibnamefont {Iwakami}}, \ and\ \bibinfo {author} {\bibfnamefont
  {K.}~\bibnamefont {Sumiyoshi}},\ }\href {\doibase
  10.1103/PhysRevD.101.023018} {\bibfield  {journal} {\bibinfo  {journal}
  {Phys. Rev. D}\ }\textbf {\bibinfo {volume} {101}},\ \bibinfo {pages}
  {023018} (\bibinfo {year} {2020})},\ \Eprint
  {http://arxiv.org/abs/1910.06176} {arXiv:1910.06176 [astro-ph.HE]}
  \BibitemShut {NoStop}%
\bibitem [{\citenamefont {Abbar}\ \emph {et~al.}(2020)\citenamefont {Abbar},
  \citenamefont {Duan}, \citenamefont {Sumiyoshi}, \citenamefont {Takiwaki},\
  and\ \citenamefont {Volpe}}]{Abbar:2019zoq}%
  \BibitemOpen
  \bibfield  {author} {\bibinfo {author} {\bibfnamefont {S.}~\bibnamefont
  {Abbar}}, \bibinfo {author} {\bibfnamefont {H.}~\bibnamefont {Duan}},
  \bibinfo {author} {\bibfnamefont {K.}~\bibnamefont {Sumiyoshi}}, \bibinfo
  {author} {\bibfnamefont {T.}~\bibnamefont {Takiwaki}}, \ and\ \bibinfo
  {author} {\bibfnamefont {M.~C.}\ \bibnamefont {Volpe}},\ }\href {\doibase
  10.1103/PhysRevD.101.043016} {\bibfield  {journal} {\bibinfo  {journal}
  {Phys. Rev. D}\ }\textbf {\bibinfo {volume} {101}},\ \bibinfo {pages}
  {043016} (\bibinfo {year} {2020})},\ \Eprint
  {http://arxiv.org/abs/1911.01983} {arXiv:1911.01983 [astro-ph.HE]}
  \BibitemShut {NoStop}%
\bibitem [{\citenamefont {Glas}\ \emph {et~al.}(2020)\citenamefont {Glas},
  \citenamefont {Janka}, \citenamefont {Capozzi}, \citenamefont {Sen},
  \citenamefont {Dasgupta}, \citenamefont {Mirizzi},\ and\ \citenamefont
  {Sigl}}]{Glas:2019ijo}%
  \BibitemOpen
  \bibfield  {author} {\bibinfo {author} {\bibfnamefont {R.}~\bibnamefont
  {Glas}}, \bibinfo {author} {\bibfnamefont {H.~T.}\ \bibnamefont {Janka}},
  \bibinfo {author} {\bibfnamefont {F.}~\bibnamefont {Capozzi}}, \bibinfo
  {author} {\bibfnamefont {M.}~\bibnamefont {Sen}}, \bibinfo {author}
  {\bibfnamefont {B.}~\bibnamefont {Dasgupta}}, \bibinfo {author}
  {\bibfnamefont {A.}~\bibnamefont {Mirizzi}}, \ and\ \bibinfo {author}
  {\bibfnamefont {G.}~\bibnamefont {Sigl}},\ }\href {\doibase
  10.1103/PhysRevD.101.063001} {\bibfield  {journal} {\bibinfo  {journal}
  {Phys. Rev. D}\ }\textbf {\bibinfo {volume} {101}},\ \bibinfo {pages}
  {063001} (\bibinfo {year} {2020})},\ \Eprint
  {http://arxiv.org/abs/1912.00274} {arXiv:1912.00274 [astro-ph.HE]}
  \BibitemShut {NoStop}%
\bibitem [{\citenamefont {Abbar}(2020)}]{Abbar:2020fcl}%
  \BibitemOpen
  \bibfield  {author} {\bibinfo {author} {\bibfnamefont {S.}~\bibnamefont
  {Abbar}},\ }\href {\doibase 10.1088/1475-7516/2020/05/027} {\bibfield
  {journal} {\bibinfo  {journal} {JCAP}\ }\textbf {\bibinfo {volume} {05}},\
  \bibinfo {pages} {027} (\bibinfo {year} {2020})},\ \Eprint
  {http://arxiv.org/abs/2003.00969} {arXiv:2003.00969 [astro-ph.HE]}
  \BibitemShut {NoStop}%
\bibitem [{\citenamefont {Johns}\ and\ \citenamefont
  {Nagakura}(2021)}]{Johns:2021taz}%
  \BibitemOpen
  \bibfield  {author} {\bibinfo {author} {\bibfnamefont {L.}~\bibnamefont
  {Johns}}\ and\ \bibinfo {author} {\bibfnamefont {H.}~\bibnamefont
  {Nagakura}},\ }\href {\doibase 10.1103/PhysRevD.103.123012} {\bibfield
  {journal} {\bibinfo  {journal} {Phys. Rev. D}\ }\textbf {\bibinfo {volume}
  {103}},\ \bibinfo {pages} {123012} (\bibinfo {year} {2021})},\ \Eprint
  {http://arxiv.org/abs/2104.04106} {arXiv:2104.04106 [hep-ph]} \BibitemShut
  {NoStop}%
\bibitem [{\citenamefont {Nagakura}\ \emph {et~al.}(2021)\citenamefont
  {Nagakura}, \citenamefont {Johns}, \citenamefont {Burrows},\ and\
  \citenamefont {Fuller}}]{Nagakura:2021hyb}%
  \BibitemOpen
  \bibfield  {author} {\bibinfo {author} {\bibfnamefont {H.}~\bibnamefont
  {Nagakura}}, \bibinfo {author} {\bibfnamefont {L.}~\bibnamefont {Johns}},
  \bibinfo {author} {\bibfnamefont {A.}~\bibnamefont {Burrows}}, \ and\
  \bibinfo {author} {\bibfnamefont {G.~M.}\ \bibnamefont {Fuller}},\
  }\href@noop {} {\  (\bibinfo {year} {2021})},\ \Eprint
  {http://arxiv.org/abs/2108.07281} {arXiv:2108.07281 [astro-ph.HE]}
  \BibitemShut {NoStop}%
\bibitem [{\citenamefont {Fuller}\ \emph {et~al.}(1987)\citenamefont {Fuller},
  \citenamefont {Mayle}, \citenamefont {Wilson},\ and\ \citenamefont
  {Schramm}}]{Fuller:1987}%
  \BibitemOpen
  \bibfield  {author} {\bibinfo {author} {\bibfnamefont {G.~M.}\ \bibnamefont
  {Fuller}}, \bibinfo {author} {\bibfnamefont {R.~W.}\ \bibnamefont {Mayle}},
  \bibinfo {author} {\bibfnamefont {J.~R.}\ \bibnamefont {Wilson}}, \ and\
  \bibinfo {author} {\bibfnamefont {D.~N.}\ \bibnamefont {Schramm}},\ }\href
  {\doibase 10.1086/165772} {\bibfield  {journal} {\bibinfo  {journal}
  {Astrophys. J.}\ }\textbf {\bibinfo {volume} {322}},\ \bibinfo {pages} {795}
  (\bibinfo {year} {1987})}\BibitemShut {NoStop}%
\bibitem [{\citenamefont {Pantaleone}(1992)}]{Pantaleone:1992eq}%
  \BibitemOpen
  \bibfield  {author} {\bibinfo {author} {\bibfnamefont {J.~T.}\ \bibnamefont
  {Pantaleone}},\ }\href {\doibase 10.1016/0370-2693(92)91887-F} {\bibfield
  {journal} {\bibinfo  {journal} {Phys.Lett.}\ }\textbf {\bibinfo {volume}
  {B287}},\ \bibinfo {pages} {128} (\bibinfo {year} {1992})}\BibitemShut
  {NoStop}%
\bibitem [{\citenamefont {Sigl}\ and\ \citenamefont
  {Raffelt}(1993)}]{Sigl:1992fn}%
  \BibitemOpen
  \bibfield  {author} {\bibinfo {author} {\bibfnamefont {G.}~\bibnamefont
  {Sigl}}\ and\ \bibinfo {author} {\bibfnamefont {G.}~\bibnamefont {Raffelt}},\
  }\href {\doibase 10.1016/0550-3213(93)90175-O} {\bibfield  {journal}
  {\bibinfo  {journal} {Nucl.Phys.}\ }\textbf {\bibinfo {volume} {B406}},\
  \bibinfo {pages} {423} (\bibinfo {year} {1993})}\BibitemShut {NoStop}%
\bibitem [{\citenamefont {Kreiss}\ and\ \citenamefont {Oliger}(1973)}]{KO3}%
  \BibitemOpen
  \bibfield  {author} {\bibinfo {author} {\bibfnamefont {H.~O.}\ \bibnamefont
  {Kreiss}}\ and\ \bibinfo {author} {\bibfnamefont {J.}~\bibnamefont
  {Oliger}},\ }\href@noop {} {\bibfield  {journal} {\bibinfo  {journal} {GARP
  publication series}\ }\textbf {\bibinfo {volume} {No. 10}},\ \bibinfo {pages}
  {Geneva} (\bibinfo {year} {1973})}\BibitemShut {NoStop}%
\bibitem [{\citenamefont {Shu}(1998)}]{Shu1998}%
  \BibitemOpen
  \bibfield  {author} {\bibinfo {author} {\bibfnamefont {C.-W.}\ \bibnamefont
  {Shu}},\ }\enquote {\bibinfo {title} {Essentially non-oscillatory and
  weighted essentially non-oscillatory schemes for hyperbolic conservation
  laws},}\ in\ \href {\doibase 10.1007/BFb0096355} {\emph {\bibinfo {booktitle}
  {Advanced Numerical Approximation of Nonlinear Hyperbolic Equations: Lectures
  given at the 2nd Session of the Centro Internazionale Matematico Estivo
  (C.I.M.E.) held in Cetraro, Italy, June 23--28, 1997}}},\ \bibinfo {editor}
  {edited by\ \bibinfo {editor} {\bibfnamefont {A.}~\bibnamefont
  {Quarteroni}}}\ (\bibinfo  {publisher} {Springer Berlin Heidelberg},\
  \bibinfo {address} {Berlin, Heidelberg},\ \bibinfo {year} {1998})\ pp.\
  \bibinfo {pages} {325--432}\BibitemShut {NoStop}%
\bibitem [{\citenamefont {Shu}(2003)}]{Shu2003}%
  \BibitemOpen
  \bibfield  {author} {\bibinfo {author} {\bibfnamefont {C.-W.}\ \bibnamefont
  {Shu}},\ }\href {\doibase 10.1080/1061856031000104851} {\bibfield  {journal}
  {\bibinfo  {journal} {International Journal of Computational Fluid Dynamics}\
  }\textbf {\bibinfo {volume} {17}},\ \bibinfo {pages} {107} (\bibinfo {year}
  {2003})},\ \Eprint
  {http://arxiv.org/abs/https://doi.org/10.1080/1061856031000104851}
  {https://doi.org/10.1080/1061856031000104851} \BibitemShut {NoStop}%
\bibitem [{\citenamefont {George}\ \emph {et~al.}(2021)\citenamefont {George},
  \citenamefont {Lin}, \citenamefont {Liu}, \citenamefont {Wu},\ and\
  \citenamefont {Xiong}}]{COSEnu}%
  \BibitemOpen
  \bibfield  {author} {\bibinfo {author} {\bibfnamefont {M.}~\bibnamefont
  {George}}, \bibinfo {author} {\bibfnamefont {C.-Y.}\ \bibnamefont {Lin}},
  \bibinfo {author} {\bibfnamefont {G.-Y.}\ \bibnamefont {Liu}}, \bibinfo
  {author} {\bibfnamefont {M.-R.}\ \bibnamefont {Wu}}, \ and\ \bibinfo {author}
  {\bibfnamefont {Z.}~\bibnamefont {Xiong}},\ }\href@noop {} {\bibfield
  {journal} {\bibinfo  {journal} {in preparation}\ } (\bibinfo {year}
  {2021})}\BibitemShut {NoStop}%
\bibitem [{\citenamefont {Banerjee}\ \emph {et~al.}(2011)\citenamefont
  {Banerjee}, \citenamefont {Dighe},\ and\ \citenamefont
  {Raffelt}}]{Banerjee:2011fj}%
  \BibitemOpen
  \bibfield  {author} {\bibinfo {author} {\bibfnamefont {A.}~\bibnamefont
  {Banerjee}}, \bibinfo {author} {\bibfnamefont {A.}~\bibnamefont {Dighe}}, \
  and\ \bibinfo {author} {\bibfnamefont {G.}~\bibnamefont {Raffelt}},\ }\href
  {\doibase 10.1103/PhysRevD.84.053013} {\bibfield  {journal} {\bibinfo
  {journal} {Phys. Rev. D}\ }\textbf {\bibinfo {volume} {84}},\ \bibinfo
  {pages} {053013} (\bibinfo {year} {2011})},\ \Eprint
  {http://arxiv.org/abs/1107.2308} {arXiv:1107.2308 [hep-ph]} \BibitemShut
  {NoStop}%
\bibitem [{\citenamefont {Capozzi}\ \emph {et~al.}(2020)\citenamefont
  {Capozzi}, \citenamefont {Chakraborty}, \citenamefont {Chakraborty},\ and\
  \citenamefont {Sen}}]{Capozzi:2020kge}%
  \BibitemOpen
  \bibfield  {author} {\bibinfo {author} {\bibfnamefont {F.}~\bibnamefont
  {Capozzi}}, \bibinfo {author} {\bibfnamefont {M.}~\bibnamefont
  {Chakraborty}}, \bibinfo {author} {\bibfnamefont {S.}~\bibnamefont
  {Chakraborty}}, \ and\ \bibinfo {author} {\bibfnamefont {M.}~\bibnamefont
  {Sen}},\ }\href {\doibase 10.1103/PhysRevLett.125.251801} {\bibfield
  {journal} {\bibinfo  {journal} {Phys. Rev. Lett.}\ }\textbf {\bibinfo
  {volume} {125}},\ \bibinfo {pages} {251801} (\bibinfo {year} {2020})},\
  \Eprint {http://arxiv.org/abs/2005.14204} {arXiv:2005.14204 [hep-ph]}
  \BibitemShut {NoStop}%
\bibitem [{\citenamefont {Shalgar}\ and\ \citenamefont
  {Tamborra}(2021{\natexlab{a}})}]{Shalgar:2020xns}%
  \BibitemOpen
  \bibfield  {author} {\bibinfo {author} {\bibfnamefont {S.}~\bibnamefont
  {Shalgar}}\ and\ \bibinfo {author} {\bibfnamefont {I.}~\bibnamefont
  {Tamborra}},\ }\href {\doibase 10.1088/1475-7516/2021/01/014} {\bibfield
  {journal} {\bibinfo  {journal} {JCAP}\ }\textbf {\bibinfo {volume} {01}},\
  \bibinfo {pages} {014} (\bibinfo {year} {2021}{\natexlab{a}})},\ \Eprint
  {http://arxiv.org/abs/2007.07926} {arXiv:2007.07926 [astro-ph.HE]}
  \BibitemShut {NoStop}%
\bibitem [{\citenamefont {Shalgar}\ and\ \citenamefont
  {Tamborra}(2021{\natexlab{b}})}]{Shalgar:2020wcx}%
  \BibitemOpen
  \bibfield  {author} {\bibinfo {author} {\bibfnamefont {S.}~\bibnamefont
  {Shalgar}}\ and\ \bibinfo {author} {\bibfnamefont {I.}~\bibnamefont
  {Tamborra}},\ }\href {\doibase 10.1103/PhysRevD.103.063002} {\bibfield
  {journal} {\bibinfo  {journal} {Phys. Rev. D}\ }\textbf {\bibinfo {volume}
  {103}},\ \bibinfo {pages} {063002} (\bibinfo {year} {2021}{\natexlab{b}})},\
  \Eprint {http://arxiv.org/abs/2011.00004} {arXiv:2011.00004 [astro-ph.HE]}
  \BibitemShut {NoStop}%
\bibitem [{\citenamefont {Martin}\ \emph {et~al.}(2021)\citenamefont {Martin},
  \citenamefont {Carlson}, \citenamefont {Cirigliano},\ and\ \citenamefont
  {Duan}}]{Martin:2021xyl}%
  \BibitemOpen
  \bibfield  {author} {\bibinfo {author} {\bibfnamefont {J.~D.}\ \bibnamefont
  {Martin}}, \bibinfo {author} {\bibfnamefont {J.}~\bibnamefont {Carlson}},
  \bibinfo {author} {\bibfnamefont {V.}~\bibnamefont {Cirigliano}}, \ and\
  \bibinfo {author} {\bibfnamefont {H.}~\bibnamefont {Duan}},\ }\href {\doibase
  10.1103/PhysRevD.103.063001} {\bibfield  {journal} {\bibinfo  {journal}
  {Phys. Rev. D}\ }\textbf {\bibinfo {volume} {103}},\ \bibinfo {pages}
  {063001} (\bibinfo {year} {2021})},\ \Eprint
  {http://arxiv.org/abs/2101.01278} {arXiv:2101.01278 [hep-ph]} \BibitemShut
  {NoStop}%
\bibitem [{\citenamefont {Rrapaj}(2020)}]{Rrapaj:2019pxz}%
  \BibitemOpen
  \bibfield  {author} {\bibinfo {author} {\bibfnamefont {E.}~\bibnamefont
  {Rrapaj}},\ }\href {\doibase 10.1103/PhysRevC.101.065805} {\bibfield
  {journal} {\bibinfo  {journal} {Phys. Rev. C}\ }\textbf {\bibinfo {volume}
  {101}},\ \bibinfo {pages} {065805} (\bibinfo {year} {2020})},\ \Eprint
  {http://arxiv.org/abs/1905.13335} {arXiv:1905.13335 [hep-ph]} \BibitemShut
  {NoStop}%
\bibitem [{\citenamefont {Cervia}\ \emph {et~al.}(2019)\citenamefont {Cervia},
  \citenamefont {Patwardhan}, \citenamefont {Balantekin}, \citenamefont
  {Coppersmith},\ and\ \citenamefont {Johnson}}]{Cervia:2019res}%
  \BibitemOpen
  \bibfield  {author} {\bibinfo {author} {\bibfnamefont {M.~J.}\ \bibnamefont
  {Cervia}}, \bibinfo {author} {\bibfnamefont {A.~V.}\ \bibnamefont
  {Patwardhan}}, \bibinfo {author} {\bibfnamefont {A.~B.}\ \bibnamefont
  {Balantekin}}, \bibinfo {author} {\bibfnamefont {t.~S.~N.}\ \bibnamefont
  {Coppersmith}}, \ and\ \bibinfo {author} {\bibfnamefont {C.~W.}\ \bibnamefont
  {Johnson}},\ }\href {\doibase 10.1103/PhysRevD.100.083001} {\bibfield
  {journal} {\bibinfo  {journal} {Phys. Rev. D}\ }\textbf {\bibinfo {volume}
  {100}},\ \bibinfo {pages} {083001} (\bibinfo {year} {2019})},\ \Eprint
  {http://arxiv.org/abs/1908.03511} {arXiv:1908.03511 [hep-ph]} \BibitemShut
  {NoStop}%
\bibitem [{\citenamefont {Roggero}(2021{\natexlab{a}})}]{Roggero:2021asb}%
  \BibitemOpen
  \bibfield  {author} {\bibinfo {author} {\bibfnamefont {A.}~\bibnamefont
  {Roggero}},\ }\href@noop {} {\  (\bibinfo {year} {2021}{\natexlab{a}})},\
  \Eprint {http://arxiv.org/abs/2102.10188} {arXiv:2102.10188 [hep-ph]}
  \BibitemShut {NoStop}%
\bibitem [{\citenamefont {Roggero}(2021{\natexlab{b}})}]{Roggero:2021fyo}%
  \BibitemOpen
  \bibfield  {author} {\bibinfo {author} {\bibfnamefont {A.}~\bibnamefont
  {Roggero}},\ }\href@noop {} {\  (\bibinfo {year} {2021}{\natexlab{b}})},\
  \Eprint {http://arxiv.org/abs/2103.11497} {arXiv:2103.11497 [hep-ph]}
  \BibitemShut {NoStop}%
\end{thebibliography}

\end{document}